\documentclass[10pt,pra,notitlepage,twocolumn,superscriptaddress,longbibliography]{revtex4-2}

\usepackage[caption=false]{subfig}
\usepackage{mathtools}
\usepackage{amsmath}
\usepackage[shortlabels]{enumitem}

\usepackage{graphicx,epic,eepic,epsfig,amsmath,latexsym,amssymb,verbatim,color}

\usepackage{amsfonts}       
\usepackage{nicefrac}       

\usepackage{amsmath}
\usepackage{bbm}

\usepackage{float}
\usepackage{tikz}
\usetikzlibrary{chains}
\usetikzlibrary{fit}
\usepackage{pgflibraryarrows}		
\usepackage{pgflibrarysnakes}		

\usepackage{epsfig}
\usetikzlibrary{shapes.symbols,patterns} 
\usepackage{pgfplots}

\usepackage[strict]{changepage}
\usepackage{hyperref}
\hypersetup{colorlinks=true,citecolor=blue,linkcolor=blue,filecolor=blue,urlcolor=blue,breaklinks=true}

\usepackage[marginal]{footmisc}
\usepackage{url}
\usepackage{theorem}

\newtheorem{definition}{Definition}
\newtheorem{proposition}{Proposition}
\newtheorem{lemma}[proposition]{Lemma}

\newtheorem{theorem}[proposition]{Theorem}

\newtheorem{corollary}[proposition]{Corollary}

\def\squareforqed{\hbox{\rlap{$\sqcap$}$\sqcup$}}
\def\qed{\ifmmode\squareforqed\else{\unskip\nobreak\hfil
\penalty50\hskip1em\null\nobreak\hfil\squareforqed
\parfillskip=0pt\finalhyphendemerits=0\endgraf}\fi}
\def\endenv{\ifmmode\;\else{\unskip\nobreak\hfil
\penalty50\hskip1em\null\nobreak\hfil\;
\parfillskip=0pt\finalhyphendemerits=0\endgraf}\fi}
\newenvironment{proof}{\noindent \textbf{{Proof~} }}{\hfill $\blacksquare$}

\newcounter{remark}

\newcounter{example}

\mathchardef\ordinarycolon\mathcode`\:
\mathcode`\:=\string"8000
\def\vcentcolon{\mathrel{\mathop\ordinarycolon}}
\begingroup \catcode`\:=\active
  \lowercase{\endgroup
  \let :\vcentcolon
  }

\usepackage{cleveref}
\usepackage{graphicx}
\usepackage{xcolor}

\RequirePackage[framemethod=default]{mdframed}
\newmdenv[skipabove=7pt,
skipbelow=7pt,
backgroundcolor=darkblue!15,
innerleftmargin=5pt,
innerrightmargin=5pt,
innertopmargin=5pt,
leftmargin=0cm,
rightmargin=0cm,
innerbottommargin=5pt,
linewidth=1pt]{tBox}

\newmdenv[skipabove=7pt,
skipbelow=7pt,
backgroundcolor=blue2!25,
innerleftmargin=5pt,
innerrightmargin=5pt,
innertopmargin=5pt,
leftmargin=0cm,
rightmargin=0cm,
innerbottommargin=5pt,
linewidth=1pt]{dBox}
\newmdenv[skipabove=7pt,
skipbelow=7pt,
backgroundcolor=darkkblue!15,
innerleftmargin=5pt,
innerrightmargin=5pt,
innertopmargin=5pt,
leftmargin=0cm,
rightmargin=0cm,
innerbottommargin=5pt,
linewidth=1pt]{sBox}
\definecolor{darkblue}{RGB}{0,76,156}
\definecolor{darkkblue}{RGB}{0,0,153}
\definecolor{blue2}{RGB}{102,178,255}
\definecolor{darkred}{RGB}{195,0,0}

\newcommand{\nc}{\newcommand}
\nc{\rnc}{\renewcommand}
\nc{\lbar}[1]{\overline{#1}}
\nc{\bra}[1]{\langle#1|}
\nc{\ket}[1]{|#1\rangle}
\nc{\ketbra}[2]{|#1\rangle\langle#2|}
\nc{\braket}[2]{\langle#1|#2\rangle}

\nc{\proj}[1]{| #1\rangle\!\langle #1 |}
\nc{\avg}[1]{\langle#1\rangle}
\nc{\rank}{\operatorname{Rank}}
\nc{\smfrac}[2]{\mbox{$\frac{#1}{#2}$}}
\nc{\tr}{\operatorname{Tr}}
\nc{\ox}{\otimes}
\nc{\dg}{\dagger}
\nc{\dn}{\downarrow}
\nc{\cA}{{\cal A}}
\nc{\cB}{{\cal B}}
\nc{\cC}{{\cal C}}
\nc{\cD}{{\cal D}}
\nc{\cE}{{\cal E}}
\nc{\cF}{{\cal F}}
\nc{\cG}{{\cal G}}
\nc{\cH}{{\cal H}}
\nc{\cI}{{\cal I}}
\nc{\cJ}{{\cal J}}
\nc{\cK}{{\cal K}}
\nc{\cL}{{\cal L}}
\nc{\cM}{{\cal M}}
\nc{\cN}{{\cal N}}
\nc{\cO}{{\cal O}}
\nc{\cP}{{\cal P}}
\nc{\cQ}{{\cal Q}}
\nc{\cR}{{\cal R}}
\nc{\cS}{{\cal S}}
\nc{\cT}{{\cal T}}
\nc{\cU}{{\cal U}}
\nc{\cV}{{\cal V}}
\nc{\cX}{{\cal X}}
\nc{\cY}{{\cal Y}}
\nc{\cZ}{{\cal Z}}
\nc{\cW}{{\cal W}}
\nc{\csupp}{{\operatorname{csupp}}}
\nc{\qsupp}{{\operatorname{qsupp}}}
\nc{\var}{{\operatorname{var}}}
\nc{\rar}{\rightarrow}
\nc{\lrar}{\longrightarrow}
\nc{\polylog}{{\operatorname{polylog}}}
\nc{\wt}{{\operatorname{wt}}}
\nc{\av}[1]{{\left\langle {#1} \right\rangle}}
\nc{\supp}{{\operatorname{supp}}}

\nc{\argmin}{{\operatorname{argmin}}}

\def\a{\alpha}

\def\i{\mathbf{i}}

\def\x{\xi}

\nc{\RR}{{{\mathbb R}}}
\nc{\CC}{{{\mathbb C}}}
\nc{\FF}{{{\mathbb F}}}
\nc{\NN}{{{\mathbb N}}}
\nc{\ZZ}{{{\mathbb Z}}}
\nc{\PP}{{{\mathbb P}}}
\nc{\QQ}{{{\mathbb Q}}}
\nc{\UU}{{{\mathbb U}}}
\nc{\EE}{{{\mathbb E}}}
\nc{\id}{{\operatorname{id}}}

\nc{\CHSH}{{\operatorname{CHSH}}}

\nc{\be}{\begin{equation}}
\nc{\ee}{{\end{equation}}}
\nc{\bea}{\begin{eqnarray}}
\nc{\eea}{\end{eqnarray}}
\nc{\<}{\langle}
\rnc{\>}{\rangle}
\nc{\rU}{\mbox{U}}

\nc{\ob}[1]{#1}

\nc{\SEP}{{\text{\rm SEP}}}
\nc{\NS}{{\text{\rm NS}}}
\nc{\LOCC}{{\text{\rm LOCC}}}
\nc{\PPT}{{\text{\rm PPT}}}
\nc{\EXT}{{\text{\rm EXT}}}
\nc{\Sym}{{\operatorname{Sym}}}

\nc{\ERLO}{{E_{\text{r,LO}}}}
\nc{\ERLOCC}{{E_{\text{r,LOCC}}}}
\nc{\ERPPT}{{E_{\text{r,PPT}}}}
\nc{\ERLOCCinfty}{{E^{\infty}_{\text{r,LOCC}}}}
\nc{\Aram}{{\operatorname{\sf A}}}

\newcommand{\eps}{\varepsilon}

\usepackage{tikz}
\usepackage{hyperref}
\hypersetup{colorlinks=true,citecolor=blue,linkcolor=blue,filecolor=blue,urlcolor=blue,breaklinks=true}

\makeatletter
\def\grd@save@target#1{%
  \def\grd@target{#1}}
\def\grd@save@start#1{%
  \def\grd@start{#1}}
\tikzset{
  grid with coordinates/.style={
    to path={%
      \pgfextra{%
        \edef\grd@@target{(\tikztotarget)}%
        \tikz@scan@one@point\grd@save@target\grd@@target\relax
        \edef\grd@@start{(\tikztostart)}%
        \tikz@scan@one@point\grd@save@start\grd@@start\relax
        \draw[minor help lines,magenta] (\tikztostart) grid (\tikztotarget);
        \draw[major help lines] (\tikztostart) grid (\tikztotarget);
        \grd@start
        \pgfmathsetmacro{\grd@xa}{\the\pgf@x/1cm}
        \pgfmathsetmacro{\grd@ya}{\the\pgf@y/1cm}
        \grd@target
        \pgfmathsetmacro{\grd@xb}{\the\pgf@x/1cm}
        \pgfmathsetmacro{\grd@yb}{\the\pgf@y/1cm}
        \pgfmathsetmacro{\grd@xc}{\grd@xa + \pgfkeysvalueof{/tikz/grid with coordinates/major step}}
        \pgfmathsetmacro{\grd@yc}{\grd@ya + \pgfkeysvalueof{/tikz/grid with coordinates/major step}}
        \foreach \x in {\grd@xa,\grd@xc,...,\grd@xb}
        \node[anchor=north] at (\x,\grd@ya) {\pgfmathprintnumber{\x}};
        \foreach \y in {\grd@ya,\grd@yc,...,\grd@yb}
        \node[anchor=east] at (\grd@xa,\y) {\pgfmathprintnumber{\y}};
      }
    }
  },
  minor help lines/.style={
    help lines,
    step=\pgfkeysvalueof{/tikz/grid with coordinates/minor step}
  },
  major help lines/.style={
    help lines,
    line width=\pgfkeysvalueof{/tikz/grid with coordinates/major line width},
    step=\pgfkeysvalueof{/tikz/grid with coordinates/major step}
  },
  grid with coordinates/.cd,
  minor step/.initial=.2,
  major step/.initial=1,
  major line width/.initial=2pt,
}
\makeatother

\usepackage{thmtools}
\usepackage{thm-restate}
\usepackage{etoolbox}
\makeatletter
\def\problem@s{}
\newcounter{problems@cnt}

\newcommand{\allproblems}{\problem@s}
\makeatother

\usepackage{amsmath,amsfonts,amssymb,array,graphicx,mathtools,multirow,bm,times,tcolorbox,relsize,booktabs}
\usepackage[utf8]{inputenc}
\usepackage[T1]{fontenc}
\usepackage{soul}
\pgfplotsset{compat=1.18}

\usepackage[ruled, vlined, linesnumbered]{algorithm2e}

\definecolor{colortwo}{rgb}{0.4,0.77,0.17}
\definecolor{colorthree}{rgb}{0.01,0.51,0.93}

\newcommand{\update}[1]{#1}

\nc{\tp}{trigonometric polynomial}
\nc{\tps}{trigonometric polynomials}
\newcommand{\UQSP}{W_{\omega,\bm\theta,\bm\phi}}
\newcommand{\UQPP}{ V_{\omega,\bm\theta,\bm\phi} }
\newcommand{\UQSPs}{W}
\newcommand{\UQPPs}{V}
\newcommand{\aux}{ {\operatorname{aux}} }

\newcommand{\mse}[1]{ {\operatorname{MSE}_{#1}} }
\newcommand{\err}[1]{ {E_{\operatorname{#1}}} }

\newcommand{\trace}[2][]{ \tr_{#1}\left[ #2 \right] }
\newcommand{\set}[1]{ \left\{ #1 \right\} }
\newcommand{\setcond}[2]{ \left\{ #1 : #2 \right\} }

\DeclarePairedDelimiter{\norm}{\lVert}{\rVert}
\DeclarePairedDelimiter{\abs}{\lvert}{\rvert}

\newcommand{\ion}{^{43}\textrm{Ca}^{+}}

\nc{\step}{\operatorname{STEP}}
\nc{\selu}{\operatorname{SELU}}
\nc{\relu}{\operatorname{ReLU}}

\allowdisplaybreaks

\begin{document}
\title{Exploring experimental limit of deep quantum signal processing using a trapped-ion simulator}

\author{J.-T.~Bu}  \thanks{Co-first authors with equal contribution}
\affiliation{Wuhan Institute of Physics and Mathematics, Innovation Academy of Precision Measurement Science and Technology, Chinese Academy of Sciences, Wuhan 430071, China}
\affiliation{University of the Chinese Academy of Sciences, Beijing 100049, China}
\author{Lei Zhang} \thanks{Co-first authors with equal contribution}
\affiliation{Thrust of Artificial Intelligence, Information Hub, The Hong Kong University of Science and Technology (Guangzhou), Guangdong 511453, China}
\author{Zhan Yu}
\affiliation{Centre for Quantum Technologies, National University of Singapore, 117543, Singapore}
\author{Jing-Bo Wang}
\affiliation{Beijing Academy of Quantum Information Sciences, Beijing 100193, China}
\author{W.-Q.~Ding}
\affiliation{Wuhan Institute of Physics and Mathematics, Innovation Academy of Precision Measurement Science and Technology, Chinese Academy of Sciences, Wuhan 430071, China}
\affiliation{University of the Chinese Academy of Sciences, Beijing 100049, China}
\author{W.-F.~Yuan}
\affiliation{Wuhan Institute of Physics and Mathematics, Innovation Academy of Precision Measurement Science and Technology, Chinese Academy of Sciences, Wuhan 430071, China}
\affiliation{University of the Chinese Academy of Sciences, Beijing 100049, China}
\author{B.~Wang}
\affiliation{Wuhan Institute of Physics and Mathematics, Innovation Academy of Precision Measurement Science and Technology, Chinese Academy of Sciences, Wuhan 430071, China}
\affiliation{University of the Chinese Academy of Sciences, Beijing 100049, China}
\author{H.-J.~Du}
\affiliation{Wuhan Institute of Physics and Mathematics, Innovation Academy of Precision Measurement Science and Technology, Chinese Academy of Sciences, Wuhan 430071, China}
\affiliation{University of the Chinese Academy of Sciences, Beijing 100049, China}
\author{W.-J.~Chen}
\affiliation{Wuhan Institute of Physics and Mathematics, Innovation Academy of Precision Measurement Science and Technology, Chinese Academy of Sciences, Wuhan 430071, China}
\affiliation{University of the Chinese Academy of Sciences, Beijing 100049, China}
\author{L.~Chen}
\affiliation{Wuhan Institute of Physics and Mathematics, Innovation Academy of Precision Measurement Science and Technology, Chinese Academy of Sciences, Wuhan 430071, China}
\affiliation{Research Center for Quantum Precision Measurement, Guangzhou Institute of Industry Co. LTD, Guangzhou, 511458, China }
\author{J.-W.~Zhang}
\affiliation{Research Center for Quantum Precision Measurement, Guangzhou Institute of Industry Co. LTD, Guangzhou, 511458, China }
\author{J.-C.~Li}
\affiliation{Guangzhou Institute of Industrial Intelligence, Guangzhou 511458, China}
\author{F.~Zhou}
\email{zhoufei@wipm.ac.cn}
\affiliation{Wuhan Institute of Physics and Mathematics, Innovation Academy of Precision Measurement Science and Technology, Chinese Academy of Sciences, Wuhan 430071, China}
\affiliation{Research Center for Quantum Precision Measurement, Guangzhou Institute of Industry Co. LTD, Guangzhou, 511458, China }
\author{Xin Wang}
\email{felixxinwang@hkust-gz.edu.cn}
\affiliation{Thrust of Artificial Intelligence, Information Hub, The Hong Kong University of Science and Technology (Guangzhou), Guangdong 511453, China}
\author{M.~Feng}
\email{mangfeng@wipm.ac.cn}
\affiliation{Wuhan Institute of Physics and Mathematics, Innovation Academy of Precision Measurement Science and Technology, Chinese Academy of Sciences, Wuhan 430071, China}
\affiliation{Research Center for Quantum Precision Measurement, Guangzhou Institute of Industry Co. LTD, Guangzhou, 511458, China }

\begin{abstract}
	Quantum signal processing (QSP), which enables systematic polynomial transformations on quantum data through sequences of qubit rotations, has emerged as a fundamental building block for quantum algorithms and data re-uploading quantum neural networks. While recent experiments have demonstrated the feasibility of shallow QSP circuits, the inherent limitations in scaling QSP to achieve complex transformations on quantum hardware remain an open and critical question. Here we report the first experimental realization of deep QSP circuits in a trapped-ion quantum simulator. By manipulating the qubit encoded in a trapped $\ion$ ion, we demonstrate high-precision simulation of some prominent functions used in quantum algorithms and machine learning, with circuit depths ranging from 15 to 360 layers and implementation time significantly longer than coherence time of the qubit. Our results reveal a crucial trade-off between the precision of function simulation and the concomitant accumulation of hardware noise, highlighting the importance of striking a balance between circuit depth and accuracy in practical QSP implementation. This work addresses a key gap in understanding the scalability and limitations of QSP-based algorithms on quantum hardware, providing valuable insights for developing quantum algorithms as well as practically realizing quantum singular value transformation and data re-uploading quantum machine learning models.
\end{abstract}

\date{\today}
\maketitle


\begin{figure*}[t]
	\centering
	\includegraphics*[width=\textwidth*9/10]{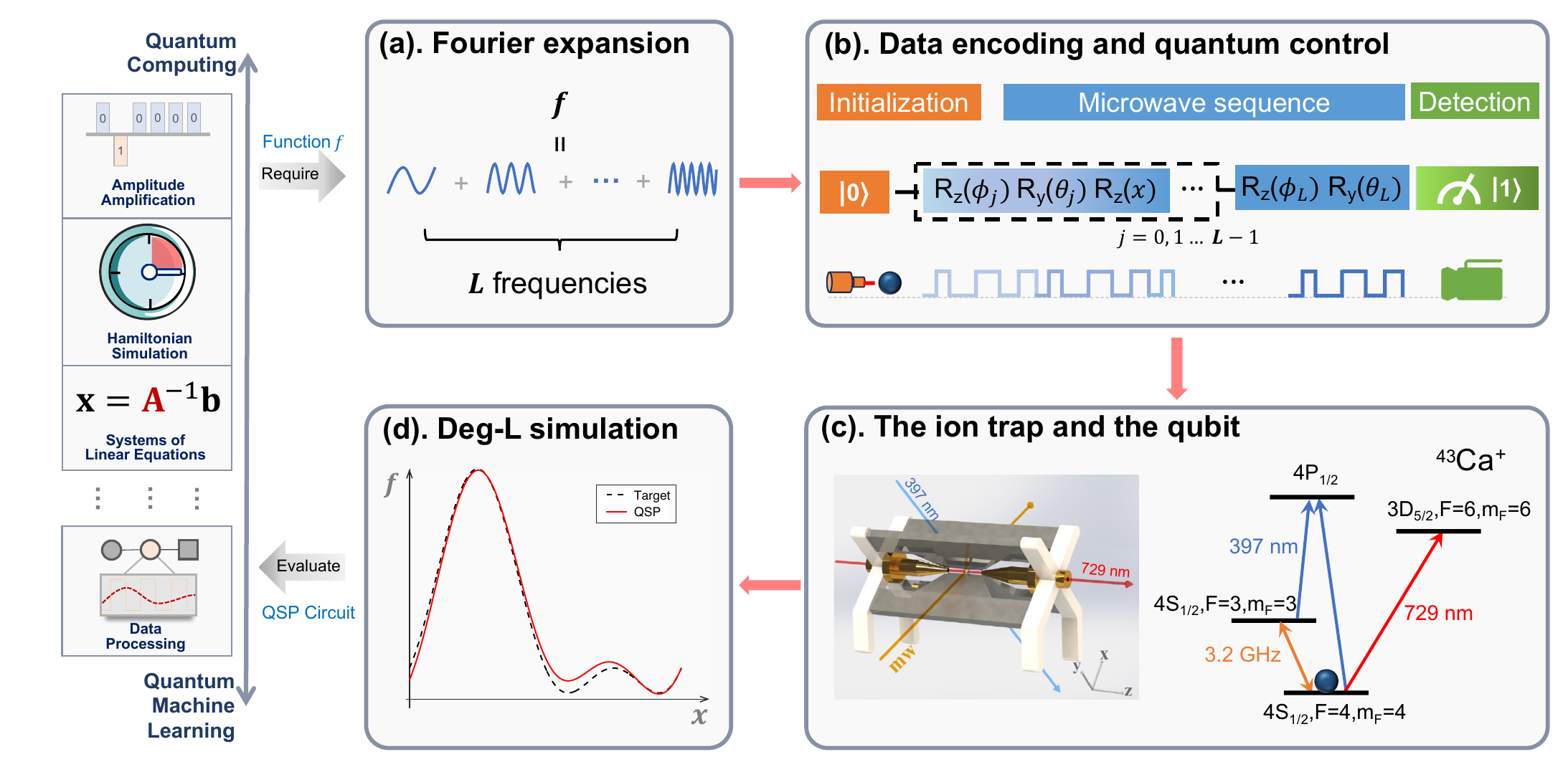}
	\caption[short] {A schematic overview of using a single trapped ion to evaluate experimental limitation of QSP. (Left panel) QSP
		finds extensive applications across various problems ranging from quantum computing to quantum machine learning tasks. The target algorithm is reduced to a task of simulating function $f$. Here it is listed as examples that amplitude amplification aligns with a linear function, Hamiltonian simulation corresponds to an evolution function, and solving linear equations requires an inverse function. Then the simulation process is divided into following four steps (\textbf{a}, \textbf{b}, \textbf{c}, \textbf{d}), which connect quantum computing to quantum machine learning. \textbf{(a)} The truncated Fourier series is computed for optimal approximation. \textbf{(b)} The QSP circuit simulating $f$ is implemented using laser and microwave sequences, from the end of the sideband cooling to the final detection. \textbf{(c)} The ion trap and the level scheme of the $\ion$ ion, where the trap's axial and radial frequencies are respectively, $\omega_z/2\pi = 1.2$ MHz and $\omega_r/2\pi  = 1.5$ MHz, and the qubit is encoded in the stretched states of the hyperfine ground state, i.e., $|0\rangle=|4S_{1/2},F=4,m_F=4\rangle$ and $|1\rangle=|4S_{1/2},F=3,m_F=3\rangle$, where F and $m_F$ represent total angular momentum and magnetic quantum number. A 3.2 GHz microwave couples the two encoded states for unitary operation. Laser cooling, optical pumping, and state measurement are performed with 397 nm and 729 nm lasers. \textbf{(d)} By analyzing the characteristic damping signals and comparing with the target function, the simulation error of this QSP circuit is incorporated back into the quantum algorithm to determine its error bounds.}
	\label{fig:main}
\end{figure*}

\section{Introduction}

Quantum computing, an emerging and pivotal field, holds immense promise for applications across various scientific domains, including cryptography~\cite{shor1994algorithms}, database searching~\cite{grover1996fast}, and machine learning~\cite{cerezo2021variational, biamonte2017quantum, larocca2022groupinvariant,QSPJ2PhysRevA_97_042315,QSPJ3PhysRevResearch_6_013241}.
By leveraging the inherent properties of quantum mechanics, such as superposition and entanglement, these algorithms can process complex computational problems, offering substantial advantages over their classical counterparts~\cite{QuantumAdvantageJ1PRXQuantum_1_020101,QuantumAdvantageJ2gibney2019hello,QuantumAdvantageJ3zhong2020quantum,QuantumAdvantageJ4PhysRevX.11.021019,QuantumAdvantageJ5Harrow2017}.

Recent advancements in quantum computing indicate that quantum signal processing (QSP)~\cite{low2016methodology} is poised to be a basic toolbox for mainstream quantum algorithms. One example is the framework of quantum singular value transformation (QSVT)~\cite{gilyen2019quantum}. By applying QSP to invariant subspaces of input data, QSVT unifies a wide range of well-established quantum algorithms~\cite{martyn2021grand}, including those designed for amplitude amplification~\cite{gilyen2019quantum}, Hamiltonian simulation~\cite{low2019hamiltonian, lloyd2021hamiltonian,martyn2023efficient,childs2018toward}, and solving of linear systems' equations~\cite{gilyen2019quantum}. This framework can be further employed to develop novel quantum algorithms that address computational tasks, such as quantum phase estimation~\cite{martyn2021grand, wang2023quantum,rall2021faster} and quantum entropy estimation~\cite{wang2023quantum,gur2021sublinear,li2019quantum,subramanian2021quantum}.

Beyond the quantum computing perspective, QSP has also emerged as a crucial tool in the field of quantum machine learning~\cite{biamonte2017quantum}. The structure of QSP is widely used in designing quantum neural networks (QNNs) to yield quantum advantage. In particular, the single-qubit data re-uploading QNN~\cite{gilvidal2020input, perez-salinas2020data} can universally approximate univariate functions~\cite{yu2022power}. This QSP-like structure has been further \update{used} to construct multi-qubit QNNs that could universally approximate multivariate functions~\cite{yu2024nonasymptotic}. Apart from directly constructing QNNs, quantum machine learning algorithms, e.g., quantum transformer~\cite{guo2024quantum}, can implement classical neural networks on a quantum computer, in which QSP is required to simulate activation functions of neural networks.

However, implementing QSP-based quantum algorithms remains challenging due to constraints from existing hardware~\cite{quantumHardwareJT1_PhysRevD_102_094505,quantumHardwareJT2_keen2021quantum,quantumHardwareJT3_PhysRevA_102_022409}, such as expensive gate cost and non-negligible noise factors~\cite{quantumNoiseJT1PRXQuantum_4_010324,quantumNoiseJT2PhysRevLett_118_140403}. As the common ground of these algorithms, it is essential to comprehend the empirical efficacy of the QSP components.
The experimental fidelity not only influences the precision of polynomial transformation in quantum algorithms, but also affects the expressivity of QSP-based QNN models. Exploiting experimental limitations~\cite{quantumlimitJT1PRXQuantum_4_027001} of the QSP will provide valuable insight into the practicality of these algorithms within the confines of quantum hardware.

The initial QSP experiment~\cite{dong2022quantum} implemented a simplified QSVT structure to evaluate quantum device performance under noises. Subsequent research works have utilized QSP circuits for Hamiltonian simulation~\cite{kikuchi2023realization} and channel discrimination~\cite{debry2023experimental} in trapped-ion systems.
Despite those achievements, the current depths of QSP circuits remain shallow. Then an open question persists: to what extent can quantum algorithms reliably handle sophisticated transformations in quantum devices?  These transformations, while significant, often exhibit singularities that cannot be approximated by low-degree polynomials. The absence of such experiments necessitates a deeper exploration for practical performance capability of QSP-based algorithms~\cite{DeepQuCirJ1PhysRevA_104_032610,DeepQuCirJ2PhysRevResearch_2_033125}.

In this work, we address this open question by investigating a tight bound for experimental errors in implementing QSP-based algorithms in the case of noise-free quantum data. Notably, our method is independent of the qubit cost of the target algorithm, \update{with} any QSP-based algorithm to be reduced to a function simulation task with only single-qubit implementation required. To showcase its effectiveness, we present the first experimental realization of deep QSP circuits in a trapped-ion quantum simulator, demonstrating the accurate execution of three important functions with singularities.
By exploiting the experimental depth of these QSP circuits, our experiments demonstrate that, within an appropriate time that far exceeds the dephasing time of the system, the average experimental error within deep QSP circuits can be exponentially suppressed.
We also identify two main sources of error and analyze their effects in deep QSP circuits. These experiments offer reliable references for assessing the fidelity of simulating deep QSP-based algorithms in ion-trap devices.



\section{Quantum signal processing}

The concept of quantum signal processing (QSP) can be traced back to the seminal work of Low et al.~\cite{low2016methodology}, who demonstrated that interleaving rotation gates can facilitate polynomial transformations of an input scalar $x$. By encoding a polynomial $P$ into the rotation angles, QSP simulates its performance $P(x)$ through the expectation value by measuring the qubit~\cite{martyn2021grand, yu2022power}. QSP can be further extended to multi-qubit frameworks~\cite{wang2023quantum, motlagh2024generalized, odake2023universal, sunderhauf2023generalized}, empowering quantum circuits to simulate polynomial transformation of input matrix data.

Many quantum algorithms can be interpreted as univariate function transformations applied to input data. QSP and its multi-qubit extension provide quantum approaches to simulate these functions and beyond. As listed in the left panel of Figure~\ref{fig:main}, each algorithm can be reduced to a task of simulating a square-integrable function $f: \RR \to \CC$. Notably, amplitude amplification corresponds to a linear function $f(x) = ax$; Hamiltonian simulation equates to an evolution function $f(x) = e^{-ixt}$; solving linear systems needs an inverse function $f(x) = x^{-1}$.
By approximating $f$ with a polynomial expansion, such as Fourier expansion~\cite{PhysRevA108032406}, one can construct a QSP-based circuit to simulate $f$ and address the corresponding problem. Nevertheless, the performance of such deep extensions on real quantum devices remains largely unexplored, while verifying a QSP-based algorithm directly would be prohibitively expensive. This raises a natural question: how effectively can we evaluate the performance of QSP-based circuits without conducting experiments that may require hundreds of qubits?

Here, we propose a method that uses only a single qubit to explore the limitations of QSP-based algorithms on NISQ devices. The approach involves first distilling the single-qubit QSP circuit from the algorithm and then carrying out the qubit-experiment. The experimental simulation error of this single-qubit circuit provides a lower bound for the experimental error in its multi-qubit extension, and such bound is tight when the signal unitary is noise-free. This method is rigorously supported by the following theorem, whose formal statement and proof are deferred to Appendix~\ref{appendix:error analysis}.

\begin{theorem}[Single-qubit QSP governs the error, informal]~\label{thm:qsp error}
    Let $\err{QSP}$ be the experimental simulation error of a single-qubit QSP circuit. Then the simulation error of its multi-qubit extension $\err{QSP-EXT}$ with noise-free signal unitary satisfies $\err{QSP-EXT} = \cO(\err{QSP})$.
\end{theorem}

In other words, when the signal unitary encoding the quantum data is noise-free, single-qubit experiments are sufficient to verify the performance of complex QSP-based algorithms. On the other hand, if the signal unitary is noisy, these experiments still provide a necessary condition for the successful implementation of these algorithms: if the single-qubit experiment fails, the multi-qubit extension will also fail.

In the rest of this paper, we will demonstrate the feasibility of our proposed method. Regarding the construction, QSP has various conventions depending on the choice of signal processing unitaries and the signal unitary~\cite{gilyen2019quantum, haah2019product, chao2020finding, yu2022power, rossi2022multivariable}. Here we consider implementing trigonometric QSP, a variant noted for its good polynomial expressiveness using a single qubit~\cite{yu2022power}. The circuit is given as
\begin{equation}
	\UQSPs(x) = A(\theta_0, \phi_0) \prod_{j=1}^{L} R_z(x) A(\theta_j, \phi_j) \label{Eq1}
\end{equation}
for $A(\theta, \phi) = R_y(\theta) R_z(\phi)$.
As shown in Fig.~\ref{fig:main}, each signal processing unitary $A(\theta_j,\phi_j)$ has been translated into a parameterized microwave sequence. Additionally, we encode the function variable $x$ into a $R_z(x)$ gate within each signal processing block using a data re-uploading process. The accuracy parameters in the signal processing unit can be trained using classical optimization methods such as gradient descend or quasi-Newton formula.

\section{Experimental system and results}

We accomplish the deep QSP circuits experimentally with a trapped $\ion$ ion in a linear Paul trap. The $\ion$ ion possesses an electronic spin-1/2 and a nuclear spin-7/2, exhibiting rich level structure. The ground-state hyperfine levels of the $\ion$ ion own a smaller inherent magnetic moment, rendering them less sensitive to magnetic field fluctuations in the environment~\cite{HighQSAccrracyJ1PhysRevX_10_011004,HighQSAccrracyJ2PhysRevLett_124_230501,HighQSAccrracyJ3niu2019universal,lucas2007longlived,lucas2014}. This results in longer coherence time, which is beneficial to enhance the performance of quantum simulator. Prior to experimental operations, we have accomplished Doppler cooling and resolved sideband cooling for the ion, yielding the final average phonon number $ \bar{n}< $ 0.1 along the axial direction with the Lamb-Dicke parameter $\eta\sim$ 0.1. Together with the optical pumping, the system is initially prepared in $|0\rangle$. Then we carry out the unitary rotations between the two encoded levels and implement projective measurement by electron shelving technique. More details are deferred to Appendix~\ref{appendix:exp}.

\begin{figure}[t]
	\centering
	\includegraphics[width=\linewidth]{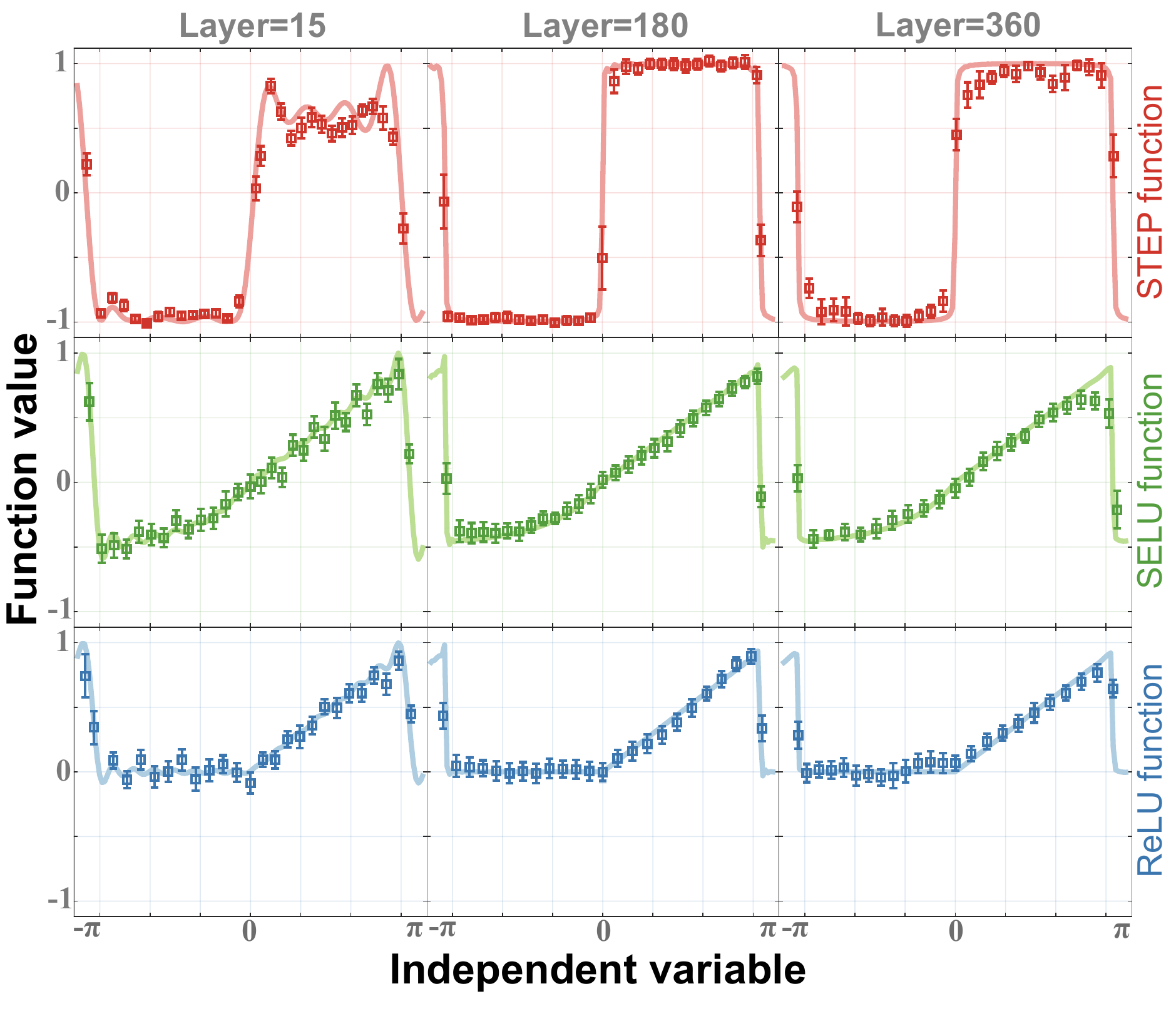}
	\caption[short]{Experimental results for simulating STEP, SELU, and ReLU functions with the layers set as $L = 15$, $180$, and $360$. The horizontal axis is the input signal ranging from $-\pi$ to $\pi$, and the vertical axis represents function values. The solid lines represent the classical simulation results, and the data points are obtained from experimental observation. Each data point is the average value of $10^5$ trials, with error bars representing the associated standard deviation.}
	\label{fig:experiment}
\end{figure}

For our purpose, the light-matter interaction is given by the Hamiltonian in units of $\hbar=1$,
\begin{equation}
	H=\frac{1}{2}\Omega(t)e^{i\varphi(t)}|e\rangle\langle g|+\rm H.C.,
	\label{Eq2}
\end{equation}
which can be achieved in the $\ion$ ion by microwave irradiation under carrier transitions. $\Omega(t)$ and $\varphi(t)$ are time dependent, representing the driving amplitude (i.e., Rabi frequency) and the phase of the microwave, respectively.
The single-qubit rotation gates $R_y(\theta)$ and $R_z(\phi)$ required by Eq. (\ref{Eq1}) can be executed using Eq. (\ref{Eq2}) by elaborately controlling $\Omega(t)$ and $\varphi(t)$.

To illustrate the experimental capabilities of QSP, we exemplify three representative functions, i.e., the STEP, SELU and ReLU for implementation.
These functions hold significant importance in the realms of quantum algorithms and machine learning.
For example, simulating the STEP function enables binary classification of input quantum data, to achieve complexity improvements on the quantum phase estimation problem~\cite{martyn2021grand, wang2023quantum, rall2021faster}.
From the machine learning perspective, the three activation functions, i.e., STEP, SELU, and ReLU, enable neural network models to become powerful function approximators~\cite{glorot2010understanding} that can satisfy the universal approximation theorem~\cite{hornik1989multilayer}.


To distinguish between inherent algorithmic limitations and hardware-induced errors, we also perform classical simulations of noise-free QSP circuits as a comparison to the experimental observation. For clarity, we refer to the experimental implementation as `quantum simulation'. Figure~\ref{fig:experiment} illustrates the behavior of quantum simulation for three representative layers with respect to the classical simulation, as below.

\emph{L=15}: Both classical and quantum simulations show limited accuracy, merely approximating the general shape of the function. The observed oscillations in function values result from an insufficient Fourier series expansion, indicating the need for increasing the layer count in order to improve the simulation fidelity.

\emph{L=180}: Simulation accuracy improves significantly with the increase of the layer count, reaching optimal performance at this configuration. However, this improvement comes at the cost of additional computational resources, with each layer requiring seven extra microwave pulses~\cite{Note1}.

\emph{L=360}: While classical simulations continue to improve in accuracy with additional layers, performance of the quantum simulation shows a marked decline. Notably, the accuracy of the quantum simulation at this setup regresses to levels comparable to those observed when $L$ ranges from 30 to 90.

\section{Error analysis}

The experimental results reveal an intriguing relationship between the experimental deviation and the circuit depth. To explore the impact of noise factors, we assess the accuracy of simulations using the mean square errors (MSEs) with respect to the ideal function, where MSEs is defined as below.
\begin{equation}
	\operatorname{MSEs}= \sum_{n=1}^{N}|W_{L}(x_{n})-f(x_{n})|^{2}/N.
	\label{Eq3}
\end{equation}
Here $W_{L}(x_{n})$ and $f(x_{n})$ are, respectively, the results of measurement (or calculation) and the ideal function. By summing the errors over $N$ variables and taking their average, the MSEs for L layers can be determined. In contrast, we treat the experimental statistical errors as the standard deviation, which reflects how much the measurement values spread out from the average.  The standard deviation is acquired by measuring the same quantity repeatedly, and labeled as error bars.
\begin{figure}[ht]
	\centering
	\includegraphics[width=\linewidth]{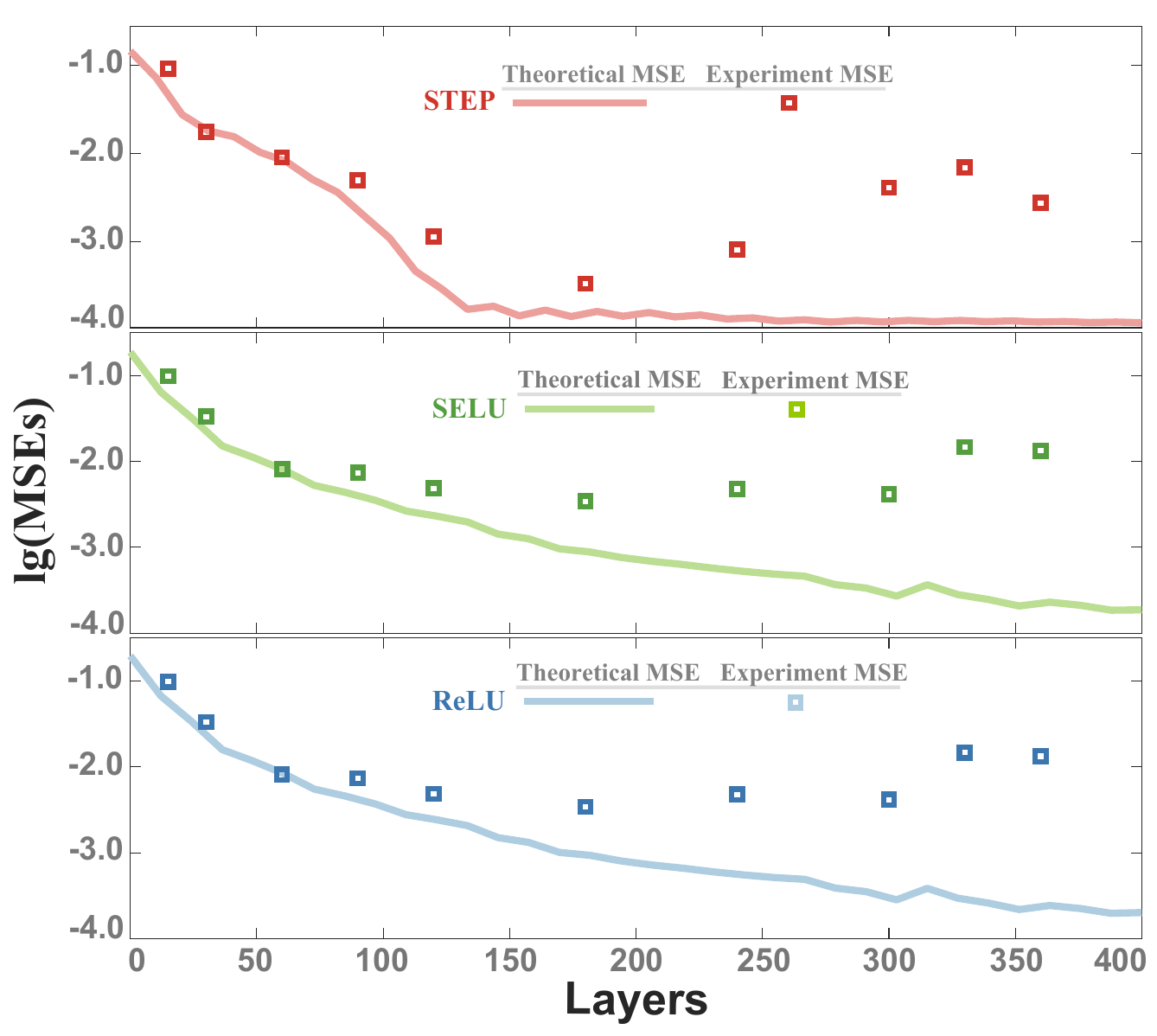}
	\caption[short]{Mean square errors (MSEs) on the logarithmic scale calculated from the results of theoretical computation and experimental observation of the QSP circuits relative to ideal function values, with layer counts ranging from 15 to 360.  Theoretical MSEs are estimated by extensive sampling of classically simulated QSP circuits, while quantum simulation MSEs are based on a limited number of experimental data points for efficiency.}
	\label{fig:MSEs}
\end{figure}
In Figure~\ref{fig:MSEs}, we evaluate quantum simulation by sampling input signals across ten different QSP circuit layer configurations. The theoretical MSEs serve as a benchmark for achievable precision, revealing an intriguing relationship between experimental MSEs and circuit depth. When the circuit is shallow, simulation accuracy is relatively low. However, since the quantum device maintains strong coherence and the accumulated errors from quantum state manipulation are minimal, the MSEs from both simulations are very close.
With the number of layers starting to increase, the experimental MSEs rapidly decrease towards zero, aligning with theoretical MSEs. This phenomenon can be well explained by the Fourier convergence theorem~\cite{weisz2012summability}. Beyond 180 layers, however, the experimental MSEs begin to increase, deviating from the expected trend and \update{reaching the minimum simulation error}. This decline stems from two primary sources, i.e., the operational errors caused by imprecise implementation, and the dephasing errors due to time-dependent decoherence. These factors create a trade-off between QSP circuit layers and simulation accuracy.

To further understand this trade-off, we classically simulate two error models based on operational and dephasing errors in the QSP circuits, tested by varying angle error rates and coherence error rates. By comparing the performance of these classically-simulated noisy QSP circuits with noise-free ones, we observe that (1) operational error is the primary source of error in the case of shallow depth, but its impact diminishes to become exponentially insignificant as depth increases to 200; (2) dephasing error is negligible in the shallow depth case but becomes exponentially significant as depth increases. These observation contributes to the kinks presented in Figure~\ref{fig:MSEs}. More discussions on these two error models are detailed in Appendix~\ref{appendix:error analysis}.

These phenomena can be explained by the nature of QSP circuits. In a QSP circuit, the $n$-th signal processing unitary governs the coefficient of the $n$-th term. Since our QSP circuits simulate polynomial expansions, the coefficients of the initial terms dominate the function value when the circuit is shallow. In this case, the operational error, affecting the signal processing unitary and thus the coefficients of expansion terms, is the main source of error. With the increase of the layer count, the significance of the affected coefficients decreases exponentially, so does the error itself. In contrast, dephasing error becomes predominant. The exponential increase in dephasing error occurs as it attempts to evolve the target state to the mixed state with reduced coherence~\cite{Note1}, counteracting the exponential decay process achieved by increasing
layers, thus increasing the MSEs exponentially.
Therefore, we consider that the two distinct noise factors exhibit contrasting behaviors as circuit depth increases, indicating a trade-off to carefully balance the circuit depth against the system's decoherence time. Such considerations will be crucial for future implementations of QSP-based algorithms and quantum neural networks.

\section{Concluding remarks}

In summary, we have proposed a method for implementing QSP-based algorithms in an ion-trap system, for which we justified that using a single-qubit one can acquire the lower bound of simulation error for complex QSP-based algorithms. Then we have experimentally carried out high-precision QSP circuits in a trapped $\ion$ ion. Utilizing quantum circuits with depths of up to 360 layers, we have executed three complex functions with singularities, confirming the potential of implementing complex QSP-based algorithms in quantum hardware. In particular, simulating the STEP function in QSVT helps unify and improve the solution of quantum estimation problems, while simulating activation functions, such as the STEP, ReLU, and SELU functions, forms a fundamental component for large-scale QNNs to achieve universal approximation. Hence, our experimental implementation of high-accuracy QSP of these functions establishes the groundwork and highlights the potential for further experimental realization of QSVT algorithms and QNNs in a trapped-ion quantum simulator.

The QSP can be extended to multi-qubit systems that encodes quantum data~\cite{camps2024explicit, sunderhauf2024blockencoding} instead of a scalar. This extension allows QSP to perform function transformations on the encoded data, thereby serving as a ground component in the design of new quantum algorithms and circuit architectures. However, the performance of such deep extensions on real quantum devices remains unexplored. With current technologies, verifying a QSP-based algorithm with signal unitaries extended to hundreds of qubits seems prohibitively expensive. Nevertheless,
the evaluation method and experiment analysis presented in this work can serve as reliable references for high-precision QSVT and data re-uploading QNN implementations, paving the way for practical applications in various domains. With further improving experimental conditions, this applications are anticipated to be constructed using deeper circuits, and hence can be leveraged to tackle challenging tasks in quantum simulation, quantum linear algebra, and quantum machine learning in the future.

\section*{Acknowledgements}

This work was supported by National Natural Science Foundation of China
under Grant Nos. 12304315, U21A20434, 92265107, 12074346, 12074390, by China
Postdoctoral Science Foundation under Grant Nos. 2022M710881, 2023T160144,
by Guangdong Provincial Quantum Science Strategic Initiative under Grant No.
GDZX2305004, by Key Lab of Guangzhou for Quantum Precision Measurement under
Grant No. 202201000010, by Science and Technology Projects in Guangzhou
under Grant Nos. 202201011727 and 2023A04J0050, and by Nansha Senior Leading
Talent Team Project under Grant No. 2021CXTD02. \
X. W. was partially supported by the National Key R\&D Program of China (Grant No.~2024YFE0102500), the National Natural Science Foundation of China (Grant. No.~12447107), the Guangdong Provincial Quantum Science Strategic Initiative (Grant No.~GDZX2403008, GDZX2403001), the Guangdong Provincial Key Lab of Integrated Communication, Sensing and Computation for Ubiquitous Internet of Things (Grant No.~2023B1212010007), the Quantum Science Center of Guangdong-Hong Kong-Macao Greater Bay Area, and the Education Bureau of Guangzhou Municipality.


\appendix

\numberwithin{equation}{section}

\section{Supplementary experimental data}~\label{appendix:extra exp data}
In this section, we present additional experimental results \update{not included} in the main text, which help further \update{understand the performance of QSP circuits}.
\begin{figure}[H]
	\centering
	\includegraphics*[width=\linewidth]{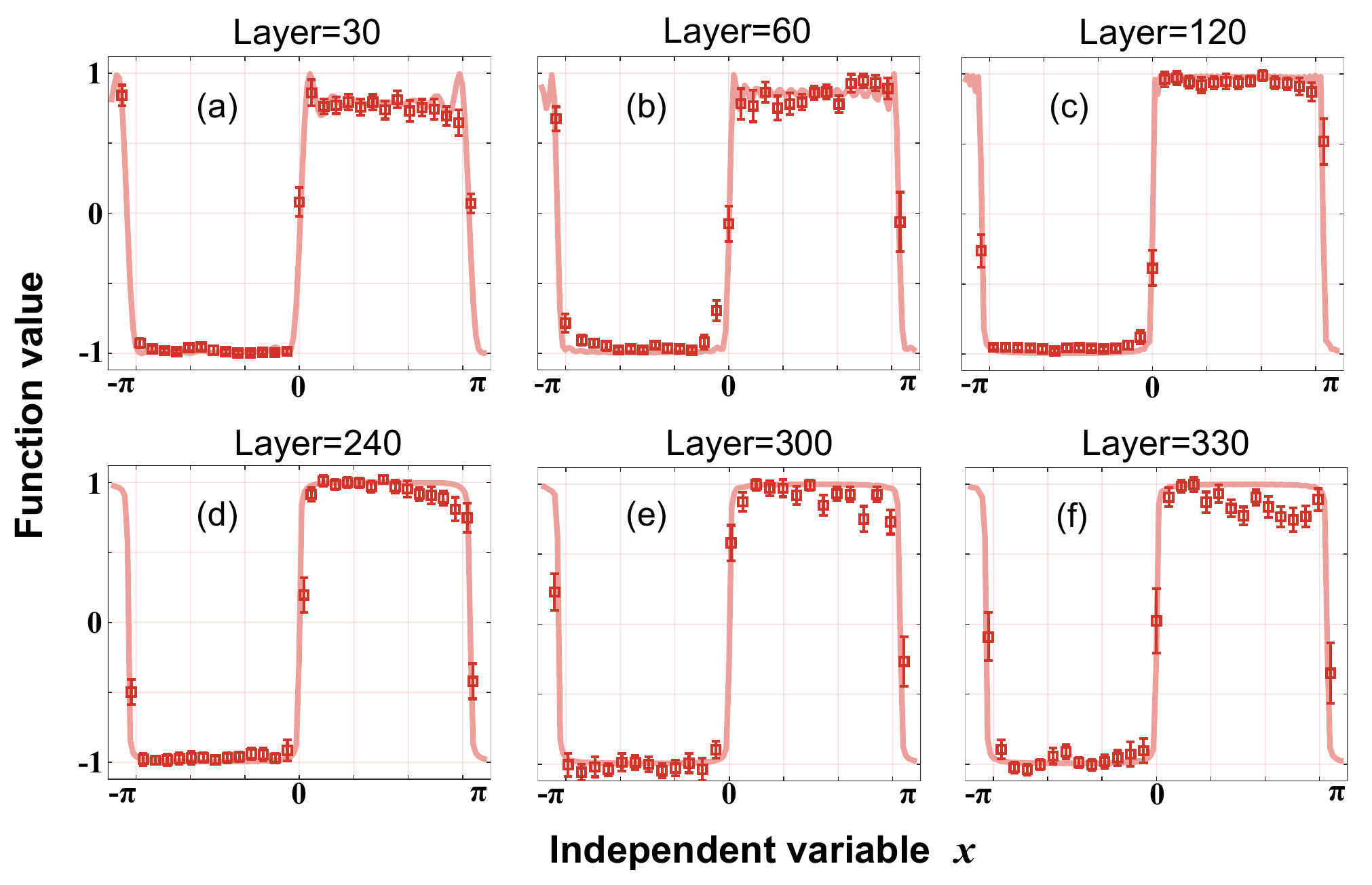}
	\caption[short]{Experimental results for the \update{STEP function}. (a)-(f) Simulation results for L=30, L=60, L=120, L=240, L=300, L=330, respectively.}
 \label{fig:FigSMStep}
 \end{figure}
\begin{figure}[H]
	\centering
	\includegraphics*[width=\linewidth]{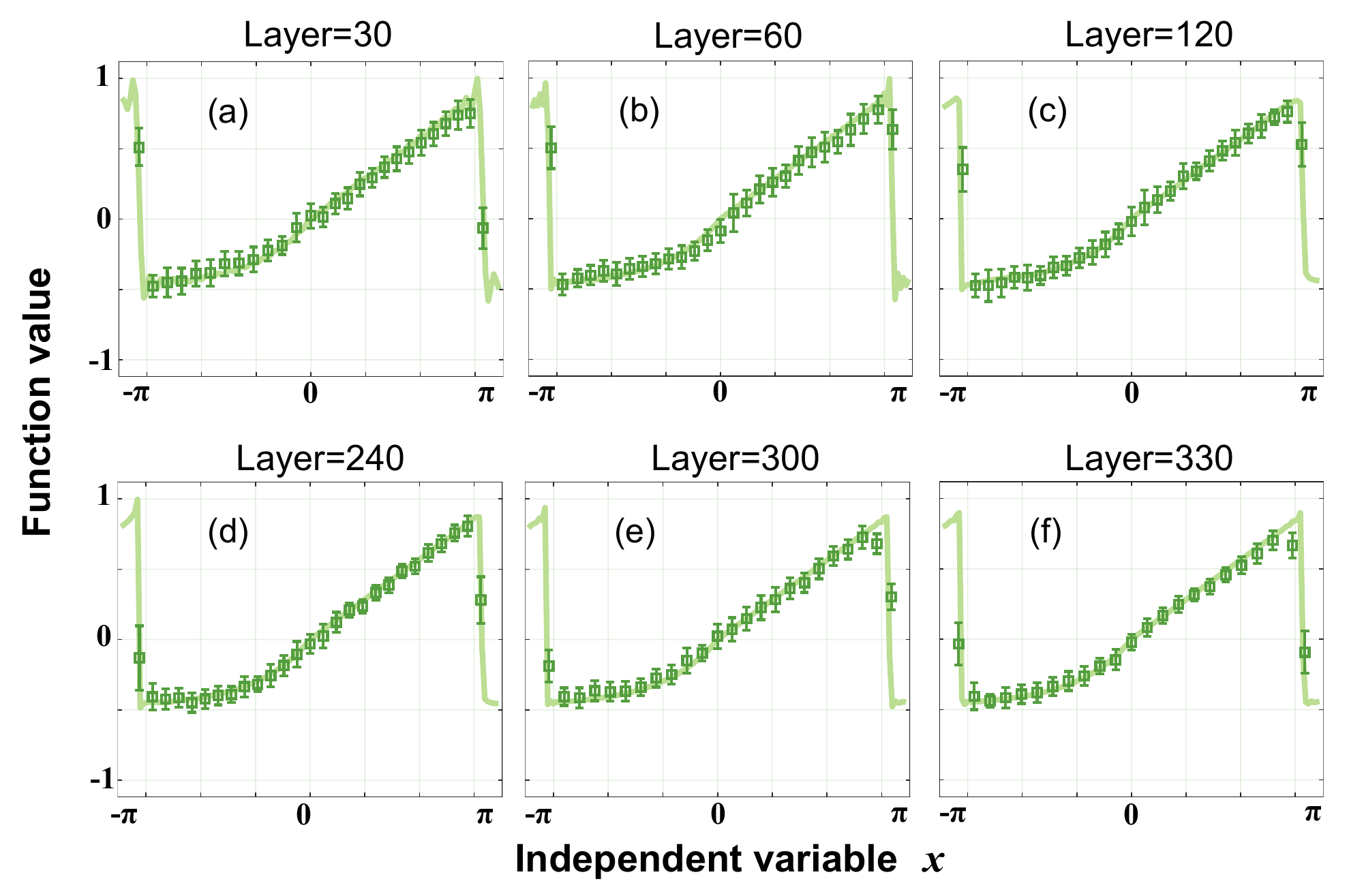}
	\caption[short]{Experimental results for the \update{SELU function}. (a)-(f) Simulation results for L=30, L=60, L=120, L=240, L=300, L=330, respectively.}
 \label{fig:FigSMSELU}
 \end{figure}
\begin{figure}[H]
	\centering
	\includegraphics*[width=\linewidth]{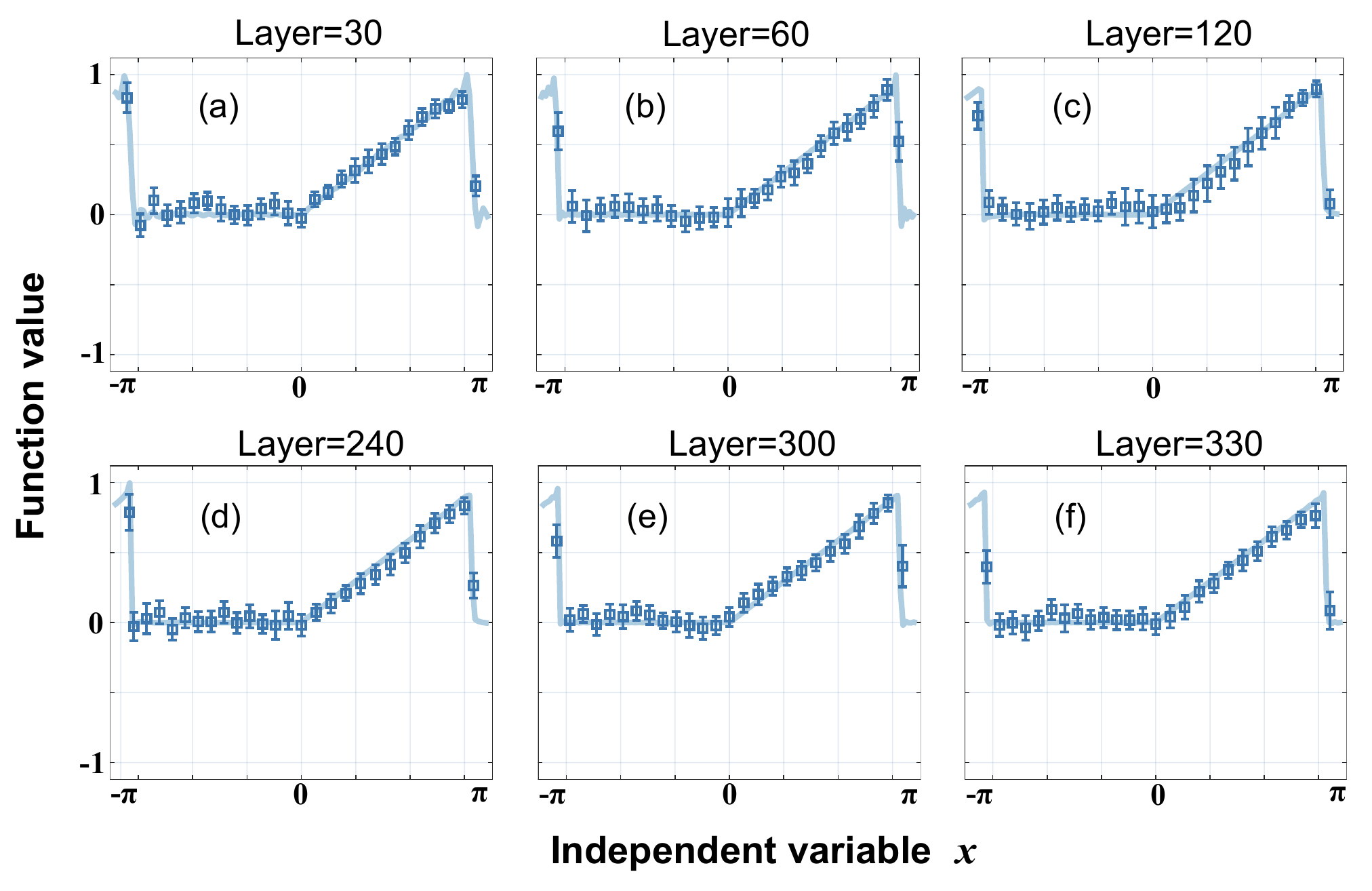}
	\caption[short]{Experimental results for the \update{ReLU function}. (a)-(f) Simulation results for L=30, L=60, L=120, L=240, L=300, L=330, respectively.}
 \label{fig:FigSMRELU}
 \end{figure}

\section{Review of Trigonometric QSP}~\label{appendix:qsp}

Our work is based on one QSP variant known as trigonometric QSP~\cite{yu2022power}, which distinguishes itself by its capability to realize general polynomial transformations using a single qubit, without resorting to other techniques such \update{as} the linear combinations of unitaries (LCU)~\cite{childs2012hamiltonian}. Construction of this QSP circuit involves a sequence of signal processing unitaries $A\left(\theta_j, \phi_j\right) = R_y(\theta_j) R_z(\phi_j)$ interleaved with a signal unitary $R_z(x)$, given as
\begin{equation}
    \UQSPs(x) = R_z(\omega) A(\theta_0, \phi_0) \prod_{j=1}^{L} R_z(x) A(\theta_j, \phi_j)
,\end{equation}
where $L$ is the degree of function approximation, and $R_z(\omega)$ \update{appearing} at the end of QSP circuit is often omitted in practical implementation. The following Lemma characterizes the power of trigonometric QSP.

\begin{lemma}[Lemma 3 in~\cite{yu2022power}]\label{lem:trig_qsp}
	There exist $\omega \in \RR$, $\bm\theta \in \RR^{L+1}$ and $\bm\phi \in \RR^{L+1}$ such that
	\begin{equation}
		\UQSP(x) = \begin{bmatrix}
			P(x) & -Q(x)\\
			Q^*(x) & P^*(x)
		\end{bmatrix},
	\end{equation}
	if and only if Laurent polynomials $P, Q \in \CC\left[e^{ix/2}, e^{-ix/2}\right]$ satisfy the following conditions,
	\begin{enumerate}
		\item $\deg(P)$, $\deg(Q) \leq L$,
		\item $P$ and $Q$ have parity $L \bmod 2$,~\label{item:parity constraint}
		\item $\forall x\in \RR$, $\abs{P(x)}^2 + \abs{Q(x)}^2 = 1$.
	\end{enumerate}
\end{lemma}

The parity condition, as delineated in Condition~\ref{item:parity constraint} and characterized by binary values of $0$ and $1$, imposes a restriction on the allowable range of polynomial transformations for QSP circuits. Generally, for a Laurent polynomial $P$ belonging to the space $\CC\left[z,z^{-1}\right]$, the parity is defined as $0$ when all the coefficients corresponding to the odd powers of $z$ are nullified, and the parity is identified as $1$ when all coefficients associated with even powers of $z$ are nullified.
This parity constraint is a commonality across various established QSP conventions, necessitating the \update{use} of the LCU technique when seeking to implement generic polynomial transformations in QSP-based algorithms. In contrast, trigonometric QSP can circumvent the parity constraint. This is achieved by restricting the scope of expressivity from $\CC\left[e^{ix/2}, e^{-ix/2}\right]$ to a smaller field $\CC\left[e^{ix}, e^{-ix}\right]$, i.e., the set of \emph{trigonometric polynomials}.
Note that \update{the definition of} the degree function in $\CC\left[e^{ix}, e^{-ix}\right]$ \update{differs} from that in $\CC\left[e^{ix/2}, e^{-ix/2}\right]$.
The implication of this restriction is summarized in the subsequent statement.

\begin{lemma}[Lemma 4 in~\cite{wang2023quantum}]~\label{lem:tri expecval}
	Suppose $F \in \CC\left[e^{ix}, e^{-ix}\right]$ is normalized and real-valued with degree $L$. There exists $P, Q \in \CC\left[e^{ix/2}, e^{-ix/2}\right]$ such that $\abs{P(x)}^2 - \abs{Q(x)}^2 = F(x)$.
	Furthermore, there exist $\omega \in \RR$, $\bm\theta \in \RR^{L+1}$ and $\bm\phi \in \RR^{L+1}$ such that $\ket{\psi(x)} = \UQSP(x) \ket{0}$ satisfies
	\begin{equation}
		\bra{\psi(x)} Z \ket{\psi(x)} = F(x)
		.\end{equation}
\end{lemma}

Using the fact that trigonometric polynomials can approximate square-integrable functions through Fourier expansion, one can eventually confirm the universal approximation properties of trigonometric QSP. When the polynomial transformation is known, we shorthand the trigonometric QSP circuit as $\UQSPs(x)$ for convenience of discussion.

\begin{theorem}[Theorem 4 in \cite{yu2022power}]
	For any \update{univariate} square-integrable function $f: [-\pi, \pi] \to [-1, 1]$ and for all $\eps > 0$, there exists a trigonometric QSP circuit $\UQSPs(x)$ such that $\ket{\psi(x)} = \UQSPs(x) \ket{0}$ satisfies
	\begin{equation}
		\norm{\bra{\psi(x)} Z \ket{\psi(x)} - f(x)} \leq \eps
		.\end{equation}
\end{theorem}

It is worth noting that when $f$ can be expressed by \update{a} convergent Fourier series, the precision will increase exponentially as the depth of $\UQSPs(x)$ increases. This theory is supported by the experiments conducted in our work.

The limits of trigonometric QSP could be representative of all types of single-qubit QSP experiments. QSP conventions can be classified into two classes: Chebyshev-based conventions~\cite{low2016methodology, gilyen2019quantum} that simulate polynomials and Fourier-based conventions~\cite{haah2019product, chao2020finding, yu2022power} that simulate trigonometric polynomials. Trigonometric QSP, as a Fourier-based convention, can express all trigonometric polynomials using a single-qubit, and can be converted to other QSP conventions \update{as shown below}
\begin{widetext}
\begin{equation}
\begin{array}{ccccc}
     \CC\left[e^{ix/2}, e^{-ix/2}\right] & \Longrightarrow & \RR\left[e^{ix/2}, e^{-ix/2}\right] & \Longleftrightarrow & \CC[x] \\
    \textrm{ trigonometric QSP } & \xrightarrow{\textrm{let } \omega, \phi_j = 0} & \textrm{ QSP in Ref.~\cite{haah2019product, chao2020finding} } & \xleftrightarrow{\textrm{add $H$ on both sides }} & \textrm{ QSP in Ref.~\cite{low2016methodology, gilyen2019quantum} }
\end{array}
\end{equation}
\end{widetext}
The details of \update{these conversions are} deferred to previous works~\cite{yu2022power, martyn2021grand}.

\subsection{QSP circuits for STEP, SELU and ReLU}~\label{appendix:example qsp}

This section illustrates how to \update{construct} the trigonometric QSP circuits for STEP, SELU and ReLU functions. First, we provide the definitions of these three functions.
Note that \update{SELU and ReLU are normalized so that} their range is restricted to $[-1, 1]$.

\begin{definition}
	For $x \in [-\pi, \pi]$, the \emph{STEP}, \emph{SELU}~\cite{klambauer2017selfnormalizing} and \emph{ReLU}~\cite{nair2010rectified} functions are defined as
	\begin{align}
		\step(x) &= \begin{cases}
			1, x \geq 0;\\
			-1, x < 0.
		\end{cases} \\
		\selu(x) &= \frac{1}{\pi} \begin{cases}
			x, x \geq 0;\\
			\a \left(e^x - 1\right), x < 0.
		\end{cases} \\
		\relu(x) &= \frac{1}{\pi} \begin{cases}
			x, x \geq 0;\\
			0, x < 0.
		\end{cases}
	\end{align}
	where $\alpha = 1.6733$ for the SELU function.
\end{definition}

The QSP circuits were computed using Paddle Quantum~\cite{paddlequantum}. Since $\selu$ and $\relu$ are continuous within the interval $(-\pi, \pi)$, these functions can be directly approximated through truncated Fourier expansion. For each function, the built-in function \emph{laurent\_generator} is utilized to determine its approximated trigonometric polynomials $F$. Subsequently, the corresponding QSP circuit in Lemma~\ref{lem:tri expecval} is constructed by employing the built-in functions \emph{pair\_generation} and \emph{qpp\_angle\_approximator}.
The simulation of $\step$ is \update{more complicated due to its jump discontinuity at $x = 0$, which exacerbates} the so-called \emph{Gibbs phenomenon}~\cite{bocher1906introduction}: the truncated Fourier expansion will exhibit heavy oscillations near the jump discontinuity. To mitigate this issue, it is observed that the STEP function can be expressed as \update{the} limit of a continuous function,
\begin{equation}
	\step(x) = \frac{2}{\pi} \lim_{N \to \infty} \arctan\left( N x \right)
	.\end{equation}
Therefore, the STEP function can be approximated by simulating $\frac{2}{\pi} \arctan\left( N x \right)$ for sufficiently large $N$ (which is chosen to be $100$ in this work).

\subsection{Multi-qubit extension}~\label{appendix:qpp}

Let $f: [-\pi, \pi] \to [-1, 1]$ be a square-integrable function. We can extend the domain of $f$ to the unitary group by applying $f$ on the eigenphases of these unitaries.
Such extension is defined as follows:

\begin{definition}[Eigenphase transformation]~\label{def:f(u)}
    Let $U$ be a unitary operator with spectral decomposition $U = \sum_j e^{i \tau_j} \ketbra{\chi_j}{\chi_j}$. The \emph{eigenphase transformation} of $U$ under $f$, denoted as $f(U)$, is defined as
\begin{equation}
    f(U) = \sum_j f(\tau_j) \ketbra{\chi_j}{\chi_j}
.\end{equation}
\end{definition}

When $f(x) = \sum_j c_j e^{ijx}$ is a trigonometric polynomial, $f(U) = \sum_j c_j U^j$ is simply a polynomial of $U$. This is where quantum phase processing (QPP)~\cite{wang2023quantum} comes into play. QPP further develops the trigonometric QSP by extending the input scalar signal to $2^n$ eigenphases of $n$-qubit unitary. By using an ancilla qubit as the control register, QPP queries the controlled input signal unitary to perform polynomial transformations on its eigenphases.
The QPP circuit for simulating degree-$L$ trigonometric function $F$ is constructed as
\begin{widetext}
\begin{equation}
    \UQPP(U) \coloneqq R_z(\omega)_\aux A\left(\theta_0, \phi_0\right)_\aux \left[
    \prod_{l=1}^{L}
    \begin{bmatrix}
        U^\dagger & 0 \\
        0 & I^{\ox n}
    \end{bmatrix} A\left(\theta_{2l - 1}, \phi_{2l - 1}\right)_\aux \begin{bmatrix}
        I^{\ox n} & 0 \\
        0 & U
    \end{bmatrix} A\left(\theta_{2l}, \phi_{2l}\right)_\aux
    \right]
,\end{equation}
\end{widetext}
where $A\left(\theta_j, \phi_j\right)_\aux$ is applied on the ancilla qubit. We shorthand the QPP circuit as $\UQPPs(U)$ for simplification.
One can show that QPP circuit essentially performs polynomial transformations on the eigenspaces of input unitary.
Lemma 2 in Ref.~\cite{wang2023quantum} shows that
\begin{equation}~\label{eqn:qpp decomp}
    \UQPPs(U) = \begin{bmatrix}
        P(U) & -Q(U)\\
        Q^*(U) & P^*(U)
    \end{bmatrix}
,\end{equation}
where $P, U$ are Laurent polynomials given in Lemma~\ref{lem:tri expecval}.

 {
\section{Detailed Error Analysis}~\label{appendix:error analysis}

The simulation error of QSP circuits for a target function $f$ can be decomposed into three components:
\begin{enumerate}[wide, labelindent=0pt]
    \item[-] \emph{polynomial approximation error} arises when $f$ is not a polynomial\update{, requiring} approximation by a truncated Fourier expansion $P$. The approximation error decrease exponentially with the degree of expansion.

    \item[-]  \emph{angle computational error} is introduced during the determination of the angle sequence in QSP circuits. The angle computation error is independent of the circuit depth, and can be reduced to the level of machine precision based on existing techniques~\cite{chao2020finding, dong2021efficient, wang2023quantum}.

    \item[-] \emph{dephasing error} is \update{a type of hardware error specific to} ion-trap devices. As the time \update{increases}, it becomes increasingly hard for devices to maintain the coherence of the qubit system, leading to more pronounced dephasing errors.

\end{enumerate}

In this section, we analyze the relation between the degree of simulation (i.e., the QSP circuit depth), and polynomial approximation error and hardware error. Before the analysis, we \update{define the} simulation error and its practical approximation.

\begin{definition}[Simulation error of single-qubit QSP]
    Let $f: [-\pi, \pi] \to [-1, 1]$ be a square-integrable function and $\hat{f}(x)$ be its approximation implemented by a QSP circuit on an ion-trap device. The \emph{simulation error} is defined as the squared $L^2$-distance between $f$ and $\hat{f}$ within $[-\pi, \pi]$,
\begin{equation}
    \err{QSP} \coloneqq \norm{f - \hat{f}}^2 = \int_{-\pi}^\pi \abs{f(x) - \hat{f}(x)}^2 \textrm{ d}x
.\end{equation}
    $\err{QSP}$ depends on the choice of $f$ and the QSP circuit, and can be approximated by sampling an ordered set of distinct points $X = \set{x_j}_{j=1}^{N}$ and computing the \emph{mean square error},
\begin{equation}
    \mse{X} = \frac{1}{N} \sum_{j=0}^{N - 1} \abs{f(x_j) - \hat{f}(x_j)}^2
.\end{equation}
\end{definition}

Note that as $N \to \infty$ and $x_{j + 1} - x_j \approx 2 \pi / N$,
\begin{equation}~\label{eqn:mse to err}
\begin{aligned}
\mse{X} &= \lim_{N \to \infty} \frac{1}{2\pi} \sum_{j=0}^{N - 1} \abs{f(x_j) - \hat{f}(x_j)}^2 (x_{j + 1} - x_j) \\
&= \frac{1}{2\pi} \err{QSP}
\end{aligned}
.\end{equation}
Therefore $\err{QSP}$ can be well approximated when $X$ is closely uniformly sampled, in which case we denote $\mse{N} = \mse{X}$ for convenience.

\subsection{Simulation error of multi-qubit extension}~\label{appendix:qpp error}

For any QPP circuit $\UQPPs(U)$ in Equation~\eqref{eqn:qpp decomp},
\begin{equation}
    \UQPPs(U) (\ket{0_\aux} \ox I) = \ket{0_\aux} \ox P(U) + \ket{1_\aux} \ox Q^*(U)
.\end{equation}
To get the eigenphase transformation under a trigonometric polynomial $F$, one need\update{s} to construct $\UQPPs(U)$ to simulate $\sqrt{ \left(1 + F(x)\right) / 2 }$, so that
\begin{equation}
    \trace[\aux]{ Z_\aux \cdot \UQPPs(U) \left(\ketbra{0_\aux}{0_\aux} \ox I \right)\UQPPs(U)^\dag } = F(U)
.\end{equation}
Similarly, one can define the simulation error of the multi-qubit extension.

\begin{definition}[Simulation error of multi-qubit QSP]
    Let $f: [-\pi, \pi] \to [-1, 1]$ be a square-integrable function and $\hat{f}(x)$ be its approximation implemented by a QSP circuit on an ion-trap device. Then the \emph{simulation error} of the multi-qubit extension of this QSP circuit is defined as the maximum performance discrepancy after applying to a quantum state,
\begin{widetext}
\begin{equation}~\label{eqn:qpp error}
    \err{QSP-EXT} = \max_{U} \setcond{ \bra{\psi} \left(f(U) - \hat{f}(U)\right)\left(f(U) - \hat{f}(U)\right)^\dagger \ket{\psi} }{ \textrm{$\ket{\psi}$ is a quantum state} }
.\end{equation}
\end{widetext}
\end{definition}

In Equation~\eqref{eqn:qpp error}, the error factors introduced by implementing $U$ or its controlled version are not considered, since these factors are dependent on $U$ and $\ket{\psi}$, and we only focus \update{on} the fundamental limitation of function simulation. In the following theorem, we show that the simulation error of the quantum eigenphase transformations under $f$ (see Definition~\ref{def:f(u)}) is equivalent to the simulation error of single-qubit QSP circuits.

\renewcommand\theproposition{\ref{thm:qsp error}}
\begin{theorem}[Single-qubit QSP governs the error, formal]
    Let $\UQSPs(x)$ be a QSP circuit that approximates a square-integrable function $f: [-\pi, \pi] \to [-1, 1]$.
    Denote $\UQPPs(U)$ as its multi-qubit extension that simulates the eigenphase transformation. Then the simulation error of the multi-qubit extension is bounded as
\begin{equation}
    \frac{1}{2\pi}\err{QSP} \leq \err{QSP-EXT} \leq \err{QSP}.
\end{equation}
\end{theorem}
\renewcommand{\theproposition}{\arabic{proposition}}
\begin{proof}
Denote $\hat{f}$
for any $n$-qubit unitary with spectral decomposition $U = \sum_j e^{i\tau_j} \ketbra{\chi_j}{\chi_j}$,
\begin{align}
    & \max_{\ket{\psi}} \bra{\psi} \left(f(U) - \hat{f}(U)\right)\left(f(U) - \hat{f}(U)\right)^\dagger \ket{\psi} \\
    =& \max_{\ket{\psi}} \sum_j \abs{f(\tau_j) - \hat{f}(\tau_j)}^2 \cdot \abs{\braket{\psi}{\chi_j}}^2 \\
    =& \max_j \abs{f(\tau_j) - \hat{f}(\tau_j)}^2
    \leq \max_{x \in [-\pi, \pi]} \abs{f(x) - \hat{f}(x)}^2\\
    =&\,\, \norm{f - \hat{f}}_\infty^2 \leq \norm{f - \hat{f}}^2 = \err{QSP}
.\end{align}
Since $U$ is arbitrary, we have $\err{QSP-EXT} \leq \err{QSP}$. For the first inequality, let $x' = \underset{x}{\mathrm{argmax}} \abs{f(x) - \hat{f}(x)}$ and choose unitary $U' =  e^{ix} I$ so that
\begin{align}
    & \max_{\ket{\psi}} \bra{\psi} \left(f(U') - \hat{f}(U')\right)\left(f(U') - \hat{f}(U')\right)^\dagger \ket{\psi} \\
    =&\,\, \abs{f(x') - \hat{f}(x')}^2 = \norm{f - \hat{f}}_\infty^2 \\
    \geq&\,\, \frac{1}{2\pi} \norm{f - \hat{f}}^2 = \frac{1}{2\pi}\err{QSP}
.\end{align}
By definition $\err{QSP-EXT} \geq \frac{1}{2\pi}\err{QSP}$.
\end{proof}

Theorem~\ref{thm:qsp error} and Equation~\eqref{eqn:mse to err} together suggest that the lower bound of the simulation error for a QSP multi-qubit extension can be estimated by experimentally computing the mean square error of its single-qubit version.

\begin{corollary}
    Let $\UQPPs(U)$ be the multi-qubit extension of a QSP circuit. Denote $\mse{N}$ as the mean square error of the QSP circuit obtained by closely uniformly sampling $N$ points in the interval $[-\pi, \pi]$. Then
\begin{equation}
    \lim_{N \to \infty} \mse{N} \leq \err{QSP-EXT}
.\end{equation}
\end{corollary}

Theorem~\ref{thm:qsp error} is also applicable to other QSP conventions. Specifically, the singular value transformation of normal matrices is defined as
\begin{equation}~\label{eqn:qsvt decomp}
    f^{\textrm{(SV)}}(A) = \sum_j f(\xi_j) \ketbra{\tilde{\psi}_j}{\psi_j}
\end{equation}
for singular value decomposition $A = \sum_j \xi_j \ketbra{\tilde{\psi}_j}{\psi_j}$.  \footnote{The basis in Equation~\eqref{eqn:qsvt decomp} could be $\ketbra{\psi_j}{\psi_j}$, depending on the choice of $f$}.
By employing a similar proof to Theorem~\ref{thm:qsp error} and using the QSP conversion discussed earlier, QSP circuits that simulate Chebyshev polynomials can similarly bound the simulation error of their multi-qubit extensions, such as the framework of quantum singular value transformation~\cite{gilyen2019quantum}.

In essence, the performance of the multi-qubit extension is fundamentally constrained by the efficacy of the single-qubit QSP circuit, even if one can implement $U$ without introducing additional noise factors. Consequently, for our objective of investigating the experimental limitations of function simulation, it is sufficient to focus on single-qubit QSP circuits in our experiments. The rationale is clear: if the single-qubit QSP circuit does not demonstrate good performance, then \update{neither does} its multi-qubit extension. Therefore, the single-qubit implementation serves as a reliable benchmark for assessing the overall capabilities of implementing QSP-based algorithms.

\subsection{Hardware error}~\label{appendix:hardware error}

In addition to the errors \update{arising} during the design of QSP circuits, executing these circuits on trapped-ion devices also results in errors due to the imperfections of the hardware, which are classified as hardware errors. In quantum simulations, our primary concerns are the errors arising from inaccuracies in experimental operations and decoherence effects.

In our experiment, we achieve various degrees of quantum state rotation by controlling the duration of microwave pulses. When the quantum simulator's control over the rotation angles lacks precision, simulation errors accumulate with increasing circuit depth, preventing the simulation accuracy from reaching the prediction by theoretical MSEs. We attribute such errors to the operational imperfection of the quantum simulator, which highlights the importance of analyzing the tolerance of QSP circuits to such errors. We define the operational error as $\operatorname{Err}=1-\{\phi, \theta, x\}_{j'}/\{\phi, \theta, x\}_{j}$, where $\{\phi, \theta, x\}_{j}$ represents the \update{ideal} parameter, and $\{\phi, \theta, x\}_{j'}$ denotes the \update{experimental} parameter. Figure \ref{fig:Op_Err_Mse} illustrates the variation of theoretical MSEs under different levels of control error. We find that even when the discrepancy between the actual and theoretical rotation angles \update{as large as} 0.1, the impact on the final simulation accuracy remains minimal. The fidelity of a single microwave $\pi$ pulse in our experiment exceeds 0.999, corresponding to Err$<$0.001. Therefore, operational errors in the experiment have a negligible effect on the final simulation accuracy.
\begin{figure}[ht]
\centering
\subfloat[]{~\label{fig:Op_Err_Mse}
    \includegraphics*[width=\linewidth]{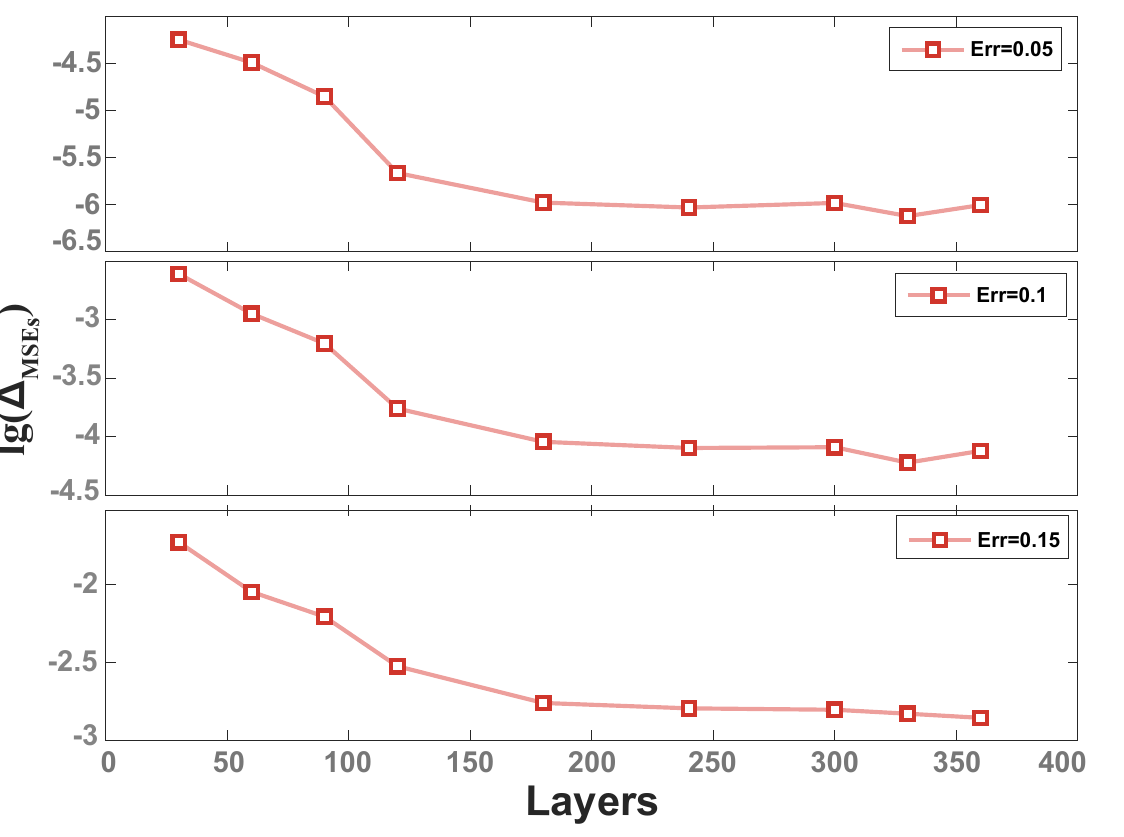}
}

\subfloat[]{~\label{fig:Dep_Err_Mse}
	\hspace{-0.5cm}\includegraphics*[width=\linewidth/10*11]{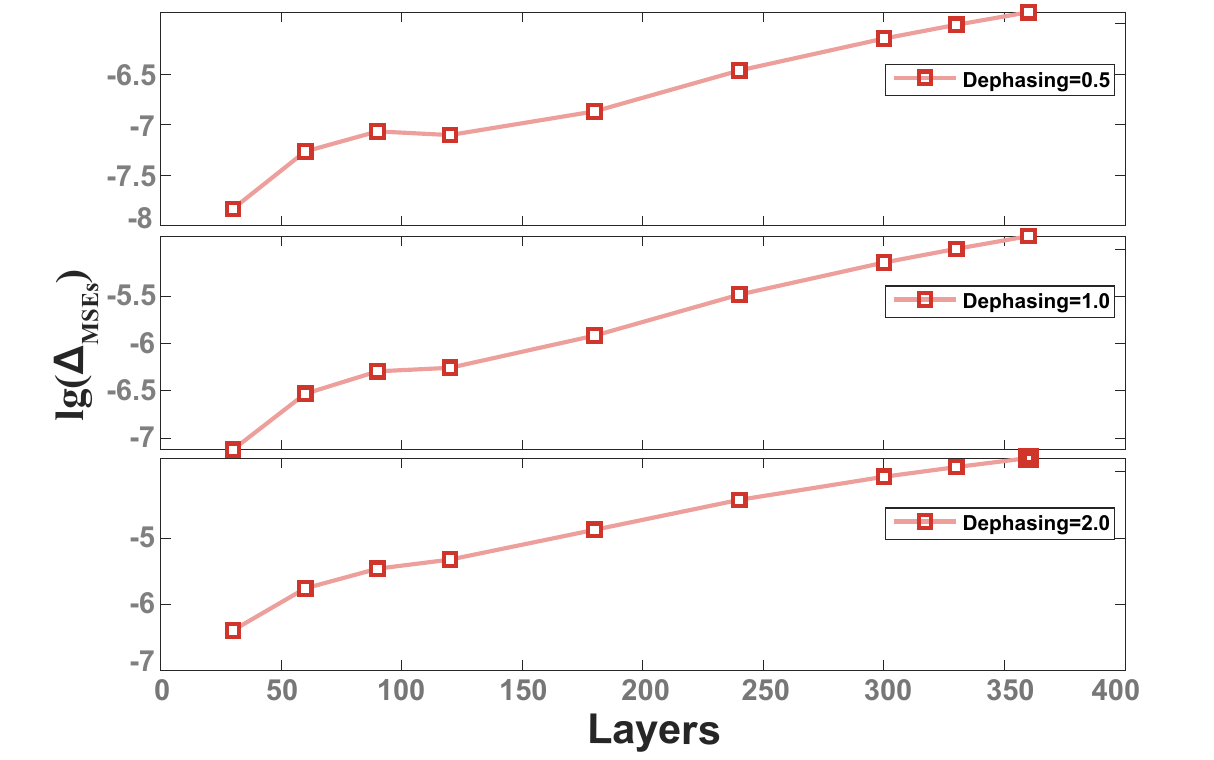}
}
\caption[short]{Six panels presenting the MSEs for nine different QSP circuit depths. \textbf{(a)} The impact of operational error on the theoretical MSEs for the STEP function. The three panels illustrate the variation in MSEs when the operational error is set to 0.05, 0.1 and 0.15 respectively. \textbf{(b)} The impact of dephasing on the theoretical MSEs for the STEP function. The three panels illustrate the variation in MSEs when the dephasing rates are set to $0.5D$, $D$ and $2D$ respectively.}~\label{fig:err_mse}
\end{figure}

To analyze the impact of decoherence on simulation accuracy, we \update{mofidy} the single qubit rotation gates $R_y(\theta_j), R_z(\phi_j), R_z(x)$ in the QSP circuit. Additionally, to study the changes in coherence during the execution of the QSP circuit, we employ the master equation to observe the evolution of the quantum system. We define the states $\ket{0}=\ket{g}= \begin{pmatrix} 0 \\ 1 \end{pmatrix}$ and $\ket{1}=\ket{e}=\begin{pmatrix} 1 \\ 0 \end{pmatrix}$, with the system initially prepared in the state $\ket{g}$, the density matrix at the initial moment is $\rho_0= |g\rangle \langle g| $. During the evolution, we introduce a decoherence term, $\frac{D}{2} \left( 2\ket{e} \langle e | \rho \ket{e} \langle e |- \ket{e} \langle e | \rho -\rho \ket{e} \langle e | \right)$, into Eq.~\eqref{Eq1} \update{to modify} $W(x)$. The decoherence rate $D \approx 2.25 \times 10^{-4} \Omega$ can be obtained from the experimentally measured Ramsey fringes, where the microwave Rabi strength $\Omega \approx 2\pi \times 35\ \update{\mathrm{kHz}}$. Figure \ref{fig:Dep_Err_Mse} shows the MSEs of the STEP function calculated at different decoherence rates.

We define the coherence of the system as the sum of the magnitudes of the off-diagonal elements of the density matrix. However, after executing the $j$-th QSP circuits, the system is not necessarily in a maximally superposed state. In our numerical simulations, we allow the system to evolve freely for one Rabi period after the $j$-th circuit. When the off-diagonal elements reach their maximum during this period, we regard the coherence of this quantum system as $C_j (\rho) = \sum_{m\neq n} \left|\rho_{m,n}\right|$.

As shown in Figure \ref{fig:Cof}, we have calculated the variation in coherence for Layer=360 of the STEP function as the system evolves. We find that the dephasing occurs more slowly during the execution of the QSP circuit than during the free evolution of the system in a noisy environment. This slower decoherence \update{heps ensure} the simulation results remain robust, even when the circuit duration exceeds the system's coherence time.
\begin{figure}[ht]
\centering
\includegraphics*[width=\linewidth]{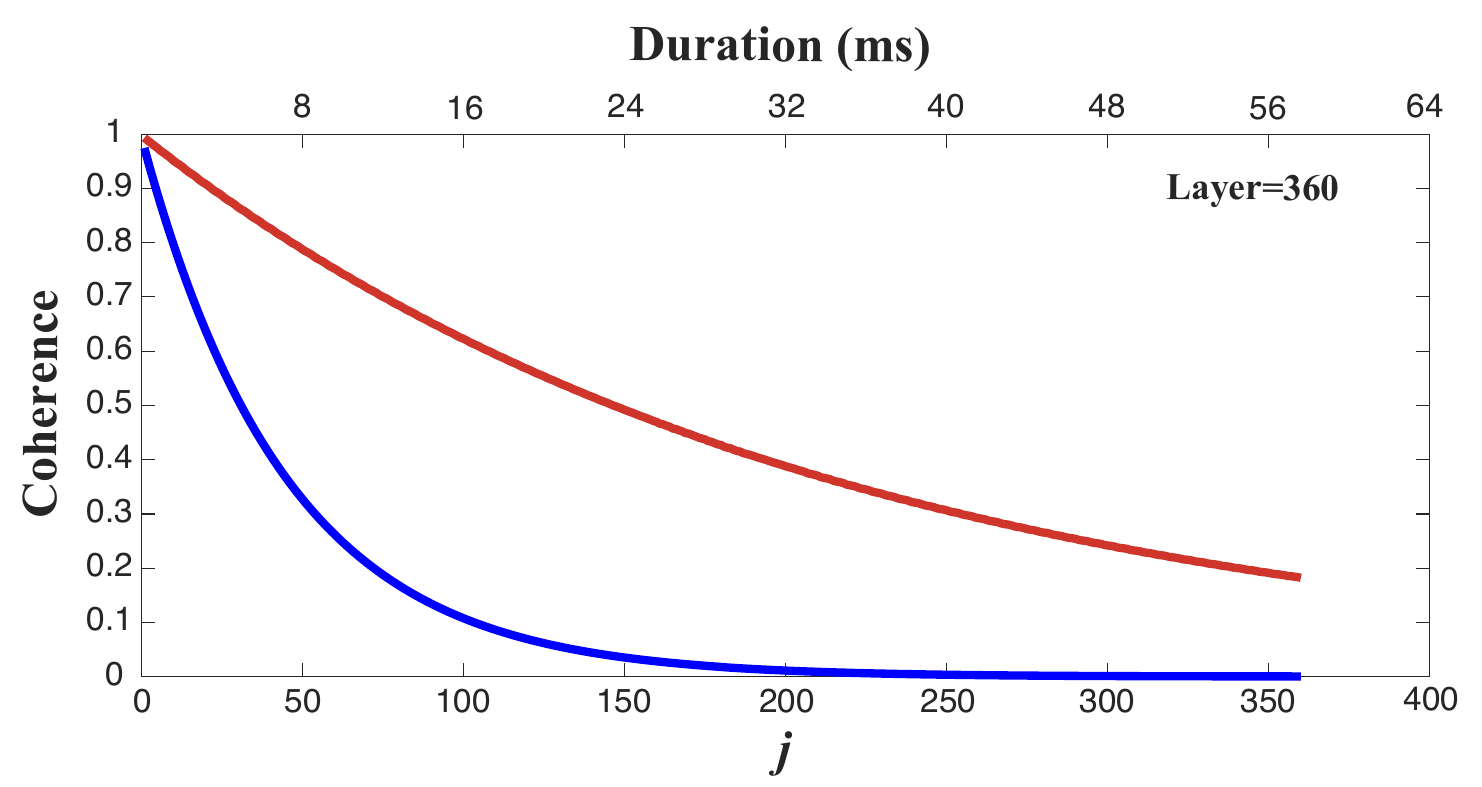}
\caption[short]{Coherence evolution for Layer=360 of the STEP function. The red and blue curves depict the coherence changes under the operation of the QSP circuits and without any operation, respectively. The horizontal axis $j$ represents the number of applied sets of $\{ \phi, \theta, x \}$, and the uppermost axis indicates the duration to reach the $j$-th circuit, and the vertical axis indicates the coherence magnitude. Each curve consists of 360 data points.}~\label{fig:Cof}
\end{figure}

\subsection{Dephasing error for a deep QSP circuit}

In Figure~\ref{fig:Dep_Err_Mse}, why would the dephasing error exponentially deteriorate the simulation precision as the layer depth decreases? One can consider a \update{greatly} simplified theoretical model of simulating a QSP circuit, by making the following assumption:

\noindent (1) $n$ is a large number and the first $n$ layers of the QSP circuit is noise-less. At this stage, the state passing the first $n$ layers can be expressed as
\begin{equation}
    \rho_n = \begin{bmatrix}
        \abs{P_n}^2 & -P_n Q_n^* \\
        -P_n^* Q_n & \abs{Q_n}^2
    \end{bmatrix}
,\end{equation}
where $P_n, Q_n$ are Laurent polynomials represented by the QSP circuit with first $L$ layers.
The hardware noise occurs right after the $n$-th layer of the QSP circuit. Further, this noise is described by a completely dephasing noise channel $\cD$, i.e., all off-diagonal entries would be removed.

The state passing the first $(n + 1)$ layers is
\begin{equation}
\begin{aligned}
    &\quad \,\,\, \rho_{n + 1}(\theta, \phi) \\
    &= \cD\left( A(\theta, \phi) \cdot \cD(\rho_{n}) \cdot A(\theta, \phi)^\dag \right) \\
    &= \cD\left( R_y(\theta) R_z(\phi) \begin{bmatrix}
        \abs{P_n}^2 & 0 \\
        0 & \abs{Q_n}^2
    \end{bmatrix} R_z^\dag(\phi) R_y^\dag(\theta) \right) \\
    &= \left[\cos^2(\theta/2)\abs{P_n}^2 + \sin^2(\theta/2) \abs{Q_n}^2\right] \ketbra{0}{0} + \\
    & \quad\, \left[\sin^2(\theta/2) \abs{P_n}^2 + \cos^2(\theta/2)\abs{Q_n}^2\right] \ketbra{1}{1}
.\end{aligned}
\end{equation}
Since $n$ is large, the $(n + 1)$-th term of the Fourier expansion of the target polynomial is small and so are $\theta$ and $\phi$. Therefore, $\rho_{n + 1}(\theta, \phi)$ evolves towards the completely mixed state and hence \update{deteriorates} the simulation result. Further, since the \update{effect is cumulative}, the rate of such evolution increase\update{s} exponentially as the layer increases.
}

\subsection{Application: quantum phase estimation}~\label{appendix:qpe}

We demonstrate how a single-qubit QSP experiment can \update{guide} the development of QSP-based algorithms to address practical problems. \update{As an example, we examine} the problem of quantum phase estimation (QPE). In this task, given a \update{unitary operator} $U$ and an eigenstate $\ket{\chi}$, the goal is to estimate its eigenphase $\tau \in [0, 2\pi)$ such that $U \ket{\chi} = e^{i \tau} \ket{\chi}$.
The \update{QPE} approach, as discussed in \cite{wang2023quantum, martyn2021grand}, employs a binary search strategy. By simulating the STEP function on the \update{unitary operator}, one can measure the ancilla qubit in the extended QSP setup to recursively classify the eigenphases, leading to the following result.

\begin{theorem}[Theorem 8 in \cite{wang2023quantum}]~\label{thm:qpe solution}
    Given a unitary $U$ and an eigenstate $\ket{\chi}$ of $U$ with eigenvalue $e^{i \tau}$, there exists a QSP-based algorithm that executes $\cO(\frac{1}{\delta} L_{\eps, \delta})$ queries to controlled-$U$ and its inverse to obtain an estimation of $\tau$ up to $\delta$ precision with probability at least $1 - \eps$. Here $L_{\eps, \delta} = \cO\left(\log(\frac{1}{\eps} \log \frac{1}{\delta}) \right)$ is the number of QSP layers.
\end{theorem}

The proof of Theorem~\ref{thm:qpe solution} intrinsically incorporates the idea of Theorem~\ref{thm:qsp error}, as detailed in Appendix C of Ref.~\cite{wang2023quantum}. For a predetermined precision $\delta$, the error probability $\eps$ of \update{the} QSP-based algorithm exhibits exponential decay as a function of increasing QSP circuit depth. This relationship can be mathematically expressed as $\eps = c_1 e^{- c_2 L_{\eps, \delta}}$, where $c_1$ and $c_2$ are constants contingent upon the input data and quantum state.

However, practical determination of an experimentally optimal $L$ to achieve a desired error rate presents a \update{non-trivial} problem, given that the input state may exist as an unknown superposition of quantum states. To address this, our numerical simulations focusing on the STEP function can give the approximate trade-off between $\eps$ and $L$. As shown in Table~\ref{tab:step error}, to achieve an error rate below $10^{-3}$, $L$ should be around 180; conversely, reducing $L$ to 30 would greatly reduce the depth of this algorithm, but make the success probability fall below 0.99.

\begin{table}[ht]
\centering
\setlength{\tabcolsep}{1em}
\caption{Detailed MSEs of simulating STEP function with respect to different depth\update{s} of deep QSP circuits.}~\label{tab:step error}
\resizebox{\linewidth}{!}{
\begin{tabular}{l|cccccccccc}
\toprule
\# of Layers & {30} & {120} & {240} & {360}\\
\midrule
\addlinespace
{experimental MSE} & {$2.8 \times 10^{-2}$} & {$2.5 \times 10^{-3}$} & {$1.9 \times 10^{-3}$} & {$5.5 \times 10^{-3}$}\\
\addlinespace
\bottomrule
\end{tabular}
}
\end{table}

Note that this trade-off is \update{independent of the input unitary operator, the specific quantum state, or the system's dimension.} Moreover, this trade-off is \update{realized} within a single qubit. Our experiments can provide a computationally efficient and resource-conservative methodology for guiding the experimental realization of QSP-based algorithm\update{s for} solving the QPE problem.

\begin{figure}[t]
\centering
	\includegraphics*[width=0.8\linewidth]{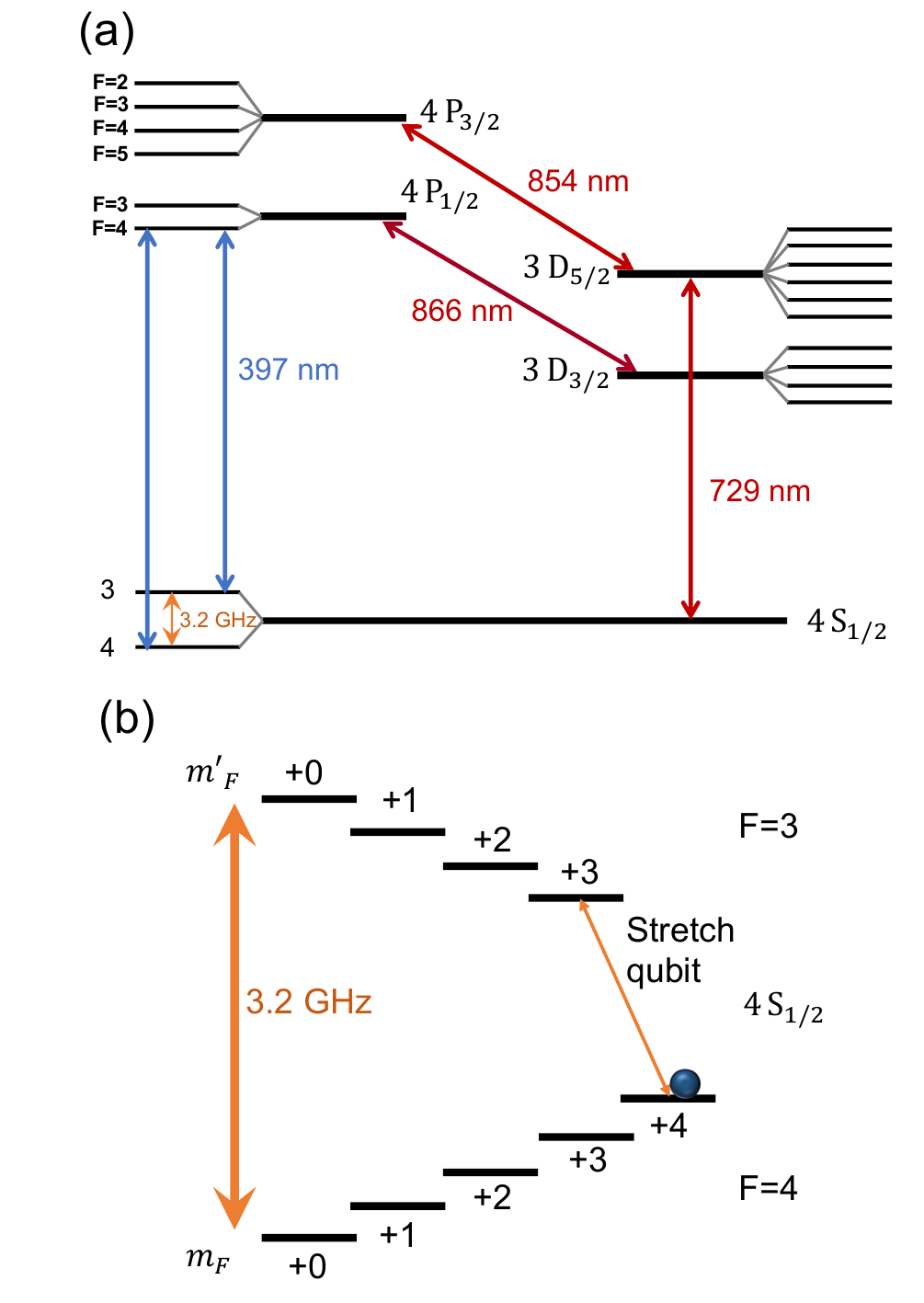}
	\caption[short]{(a) The level scheme and hyperfine splitting of the lowest energy levels of the  $\ion$ ion. Two lasers at 397 nm are used for Doppler cooling and detection. Additionally, lasers at 854 nm and 866 nm serve to repump from the D states to the P states. The ultra stable laser at 729 nm is used for sideband cooling and optical shelving. (b) The ground level hyperfine structure in an external magnetic field. Microwave at 3.2 GHz is used for the ground state transition. The qubit is encoded to Stretch transition.}
\end{figure}

\section{Experiment details}~\label{appendix:exp}

\subsection{Setup and laser cooling}~\label{appendix:exp setup}
Our experiment employs a linear Paul trap. The four blade electrodes carry the rf potential for radial confinement and the end-cap electrodes with dc voltage provide confinement in the axial direction. The trap is driven at $\Omega_{rf}/2\pi=23.9$ MHz with the power of 3.5 \update{W}. The trap axial frequency is $\omega_{z}/2\pi=1.2$ MHz with a voltage of 600 V applied to the end-cap electrodes.
A magnetostatic field of 8.2 \update{G}, generated by a pair of permanent magnets, defines a quantization axis and splits the energy levels of the ground state $4S_{1/2}$, the metastable state $3D_{5/2}$, and the excited state $4P_{1/2}$ into 16, 48 and 16 hyperfine energy levels respectively. The laser and microwave scheme can be found in Fig. \ref{fig:exp pulse}, where an ultra-stable narrow linewidth optical fiber laser for 729 nm transitions and the \update{3.2 GHz microwave are critical for the} implementation of the experiment. The 729 nm fiber laser is locked to a high-finesse cavity made of ultra-low expansion material.  A typical linewidth (FWHM) of 10 Hz is measured from the heterodyne beat note with respect to another laser system. The long-term drift of the 729 nm laser is measured to be 0.04 Hz/s by observing transitions in the ion. The microwave is directed and amplified by a horn antenna. We manipulate the qubit by controlling the frequency, amplitude, and phase of microwave by the direct digital synthesizer (DDS) and field programable gate array (FPGA) as shown in section IV.

For Doppler cooling and fluorescence detection, the ion is
excited on the $4S_{1/2} \leftrightarrow 4P_{1/2}$ dipole transition by two 397 nm lasers with a frequency difference of approximately $3.2$ GHz. To avoid optical pumping into the $3D_{3/2}$ manifold, the 866 nm repumping laser has to be applied. We enhance the repumping efficiency by finely tuning the laser to the $|3D_{3/2},F=3\rangle\leftrightarrow |4P_{1/2},F=3\rangle$ transition frequency and red-shifting a portion of the light using two acousto-optical modulators operating at frequencies of 145 MHz and 235 MHz. In this manner, all hyperfine $3D_{3/2}$ levels are resonantly coupled to one of the $|4P_{1/2},F=3,4\rangle$ levels.
We employ two 397 nm right-handed circularly polarized lasers to prepare the initial state $|4S_{1/2},F=4,m_f=4\rangle$. Subsequently, we employ 729 nm laser to shelve the population of $|4S_{1/2},F=4,m_F=4\rangle$ to $|3D_{5/2}, F=6, m_F=6\rangle$ by a $\pi$ pulse. Then we can use this quadrupole transition  for quantum state detection by electron shelving method. High-fidelity qubit operations require the ion \update{to be cooled within} the Lamb-Dicke regime and close to the motional ground state. This is achieved by applying sideband cooling on the narrow  $|4S_{1/2},F=4,m_F=4\rangle$ to $|3D_{5/2}, F=6, m_F=6\rangle$ transition. Since the meta-stable $D_{5/2}$ level is long lived, it needs to be quenched by the 854 nm \update{repumping} laser. For optimal performance, the repumping laser intensity has to be modified as the ion is cooled to the ground vibrational state.

\subsection{Experimental pulse sequences}~\label{appendix:pulse}
\update{Microwave frequency sources} are based on the frequency mixing of the commercial signal generator and DDS. The use of the DDS allows for phase and frequency control during a single experimental sequence. $R_y(\theta_j), R_z(\phi_j), R_z(x)$ denote rotations of the quantum state around different axes of the Bloch sphere by varying angles. optimized parameters $\theta_j, \phi_j$ are converted into microwave pulse sequences with different pulse time and phases.

\begin{figure}[ht]
	\centering
	\includegraphics*[width=\linewidth]{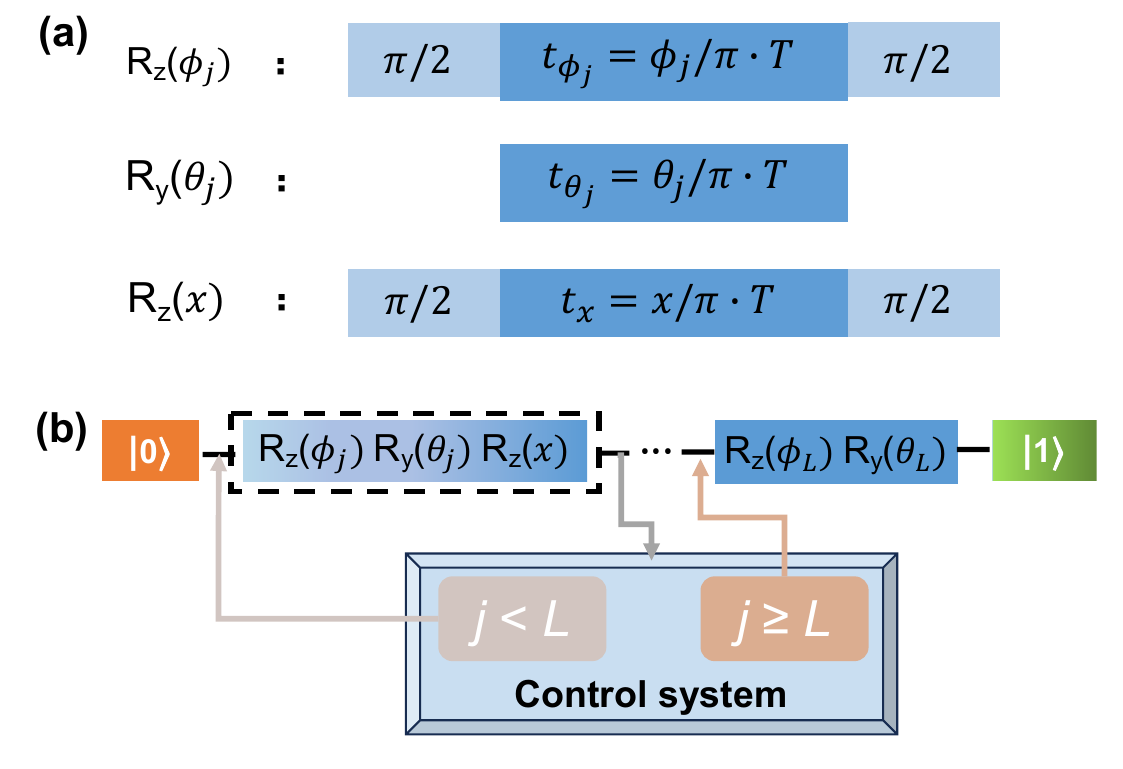}
	\caption[short]{Experimental pulse sequences and experimental scheme for QSP. (a) is the schematic diagram of microwave pulse sequences for implementing $R_y(\theta_j), R_z(\phi_j)$ and $R_z(x)$  rotations. $\pi / 2$ in the light colored box represents the microwave $\pi / 2$ pulse. $t_{\phi_j},t_{\theta_j}$ and $t_{x}$ in the dark box denote the duration of the $R_z(\phi_j),R_y(\theta_j)$ and $ R_z(x)$ rotations, respectively. (b) is the schematic diagram of implementing the QSP circuit. When the loop count $j$ is less than L, experimental control system converts the optimized parameters $\phi_j,\theta_j$ into microwave pulse sequences. When the loop count $j$ exceeds L, the QSP circuit stops working.}
 \label{fig:exp pulse}
\end{figure}

In Fig. \ref{fig:exp pulse}(a), a microwave pulse sequence \update{is generated with an} initial phase of $\frac{\pi}{2}$ and duration of $t_{\theta_j}$ \update{is generated} to implement $R_y(\theta_j)$, where $R_y(\theta_j)$ signifies the evolution of the quantum state around $y$ axis of the Bloch sphere for the angle of $\theta_j$, $R_z(\phi_j)$ is the evolution of quantum state along $z$ axis of the Bloch sphere with the angle of $\phi_j$. The operations can be achieved by controlling FPGA to output three microwave pulses. The first pulse, with initial phase of $\pi/2$ and duration of $T/2$ ($T$ is the duration of the $\pi$ pulse), rotates the quantum state in the equatorial plane of the Bloch sphere. The second pulse, with initial phase of 0 and duration of $t_{\phi_j}$, induces an evolution around  $z$ axis of the Bloch sphere with the angle of $\phi_j$, where $t_{\phi_j}=\phi_jT /\pi$. The third pulse is similar to the first one, rotating the quantum state back from the equatorial plane to its original axis. Both $R_z(x)$ and $R_z(\phi_j)$ involve similar operations.
Experimentally, for simulation layers $L$, the control system continuously outputs microwave pulses to obtain function value $\UQSPs(x)$ in the trapped-ion system, following the quantum circuit given in Eq. (\ref{Eq1}).
\update{Increasing} $L$ by one necessitates \update{seven additional} microwave pulses. Therefore, it can be simplified by adding a loop to enable the control system to automatically output microwave pulses. As shown in Fig. \ref{fig:exp pulse}(b), we prepare the initial state of $\ion$ ion in $\ket{0}$. Subsequently, the control system \update{determines} whether the loop count is \update{greater} than $L$. If not, it converts the optimized parameters $\theta_j, \phi_j$ and the independent variable $x$ into the $R_y(\theta_j), R_z(\phi_j)$ and $ R_z(x)$ rotations, respectively. Otherwise, the loop terminates and we make measurements of the states of the ion.

\subsection{Measurement of coherence}~\label{appendix:measurement}
\begin{figure}[t]
	\centering
	\includegraphics*[width=\linewidth]{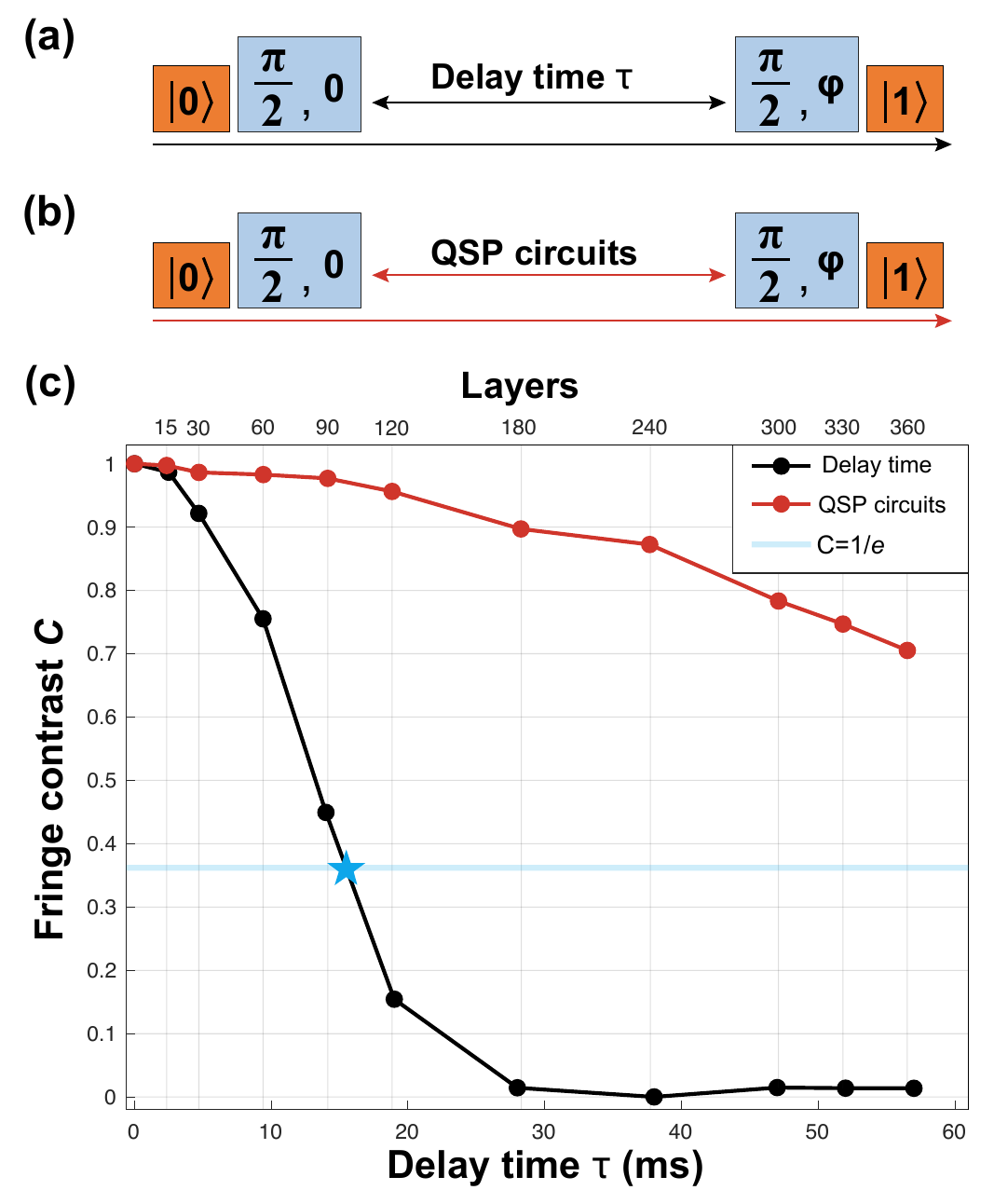}
	\caption[short]{Ramsey phase experiment on the stretched state transition.  Between the two $\pi/2$ pulses, the delay time $\tau$ and QSP circuits are inserted in (a) and (b), respectively. At each value of the Ramsey free precession time  $\tau$, the phase of the second $\pi/2$ pulse is varied to produce a set of Ramsey fringes. (c) Fringe contrast of (a) and (b), where the black points represent the fringe contrast obtained by inserting different delay time $\tau$ in (a), with $\tau=0, 2.5, 4.7, 9.4, 14, 19, 28, 38, 47, 52, 57$ ms, respectively. The red points represent the fringe contrast obtained by inserting QSP circuits with different layers of the STEP function in (b), with layers to be $0, 15, 30, 60, 90, 120, 180, 240, 300, 330, 360$, respectively. The blue horizontal line represents the case where the fringe contrast equals to $1/e$. The blue star indicate the coherence time of 16.5 ms measured by (a).}
 \label{fig:Ramsey}
\end{figure}

To measure the coherence time of the qubit, a Ramsey experiment \update{is performed} on the stretched state transition, as shown in Fig. \ref{fig:Ramsey}(a). The phase of the first $\pi/2$ pulse \update{is set} to 0, and then the phase $\varphi$ of the second $\pi/2$ pulse \update{is scanned} after a delay time.
For a fixed delay time $\tau$, the population of $\ket{1}$ \update{oscillates} as the $\pi/2$ pulse changes. The fringe contrast is $C=(P_{max}-P_{min})/(P_{max}+P_{min})$,
where $P_{max}$ and $P_{min}$ are, respectively, the maximum and minimum values of the population in this oscillation. When the fringe contrast reaches its maximum value ($C=1$ with $P_{max}=1$ and $P_{min}=0$), it indicates the maximum coherence. When the fringe contrast reaches its minimum value ($C=0$ with $P_{max}=0.5$ and $P_{min}=0.5$), it indicates no coherence.
In Fig. \ref{fig:Ramsey}(b), by substituting the delay time with QSP circuits of different layers, we can assess the impact of QSP on coherence. To \update{analyze} the impact of QSP circuits on coherence, we set the delay time to be equivalent to the duration of the microwave pulse sequences for different layers. In Fig. \ref{fig:Ramsey}(c), we find that the coherence of the system is significantly improved after the QSP circuit is executed. We conjecture that the component pulses in QSP \update{partially act as a spin echo}, so that
the simulation accuracy remains high even when the duration of the QSP circuits far exceeds the coherence time of the system.


\bibliography{references}

\end{document}